\documentclass[twocolumn, switch]{article} 

\usepackage{preprint}

\usepackage{amsmath, amsthm, amssymb, amsfonts}

\usepackage[numbers,square]{natbib}
\usepackage[utf8]{inputenc}	
\usepackage[T1]{fontenc}	
\usepackage{xcolor}		
\usepackage[colorlinks = true,
            linkcolor = purple,
            urlcolor  = blue,
            citecolor = cyan,
            anchorcolor = black]{hyperref}	
\usepackage{booktabs} 		
\usepackage{nicefrac}		
\usepackage{microtype}		
\usepackage{lineno}		
\usepackage{float}			

\usepackage{changepage}  
\usepackage{caption}          

\usepackage{newfloat}
\DeclareFloatingEnvironment[name={Supplementary Figure}]{suppfigure}
\usepackage{sidecap}
\sidecaptionvpos{figure}{c}

\usepackage{titlesec}
\titlespacing\section{0pt}{12pt plus 3pt minus 3pt}{1pt plus 1pt minus 1pt}
\titlespacing\subsection{0pt}{10pt plus 3pt minus 3pt}{1pt plus 1pt minus 1pt}
\titlespacing\subsubsection{0pt}{8pt plus 3pt minus 3pt}{1pt plus 1pt minus 1pt}

\usepackage{tikz,xcolor,hyperref}

\definecolor{lime}{HTML}{A6CE39}
\DeclareRobustCommand{\orcidicon}{
	\begin{tikzpicture}
	\draw[lime, fill=lime] (0,0)
	circle [radius=0.16]
	node[white] {{\fontfamily{qag}\selectfont \tiny ID}};
	\draw[white, fill=white] (-0.0625,0.095)
	circle [radius=0.007];
	\end{tikzpicture}
	\hspace{-2mm}
}
\foreach \x in {A, ..., Z}{\expandafter\xdef\csname orcid\x\endcsname{\noexpand\href{https://orcid.org/\csname orcidauthor\x\endcsname}
			{\noexpand\orcidicon}}
}

\title{Improved rainfall forecasts in daily use over East Africa}

\usepackage{xwatermark}
\newwatermark[firstpage,color=gray!90,angle=0,scale=0.28, xpos=0in,ypos=-5in]{*correspondence: \texttt{fenwick.cooper@physics.ox.ac.uk}}

\usepackage{authblk}

\author[a,*]{Fenwick C. Cooper}
\author[a]{Shruti Nath}
\author[a]{Andrew T. T. McRae}
\author[a]{Bobby Antonio}
\author[a]{Matthew Wright}
\author[a,b,f]{Antje Weisheimer}
\author[a]{Tim Palmer}
\author[c]{Masilin Gudoshava}
\author[c]{Nishadh Kalladath}
\author[c]{Ahmed Amidhun}
\author[c]{Jason Kinyua}
\author[d]{Hannah Kimani}
\author[d]{David Koros}
\author[d]{Zacharia Mwai}
\author[d]{Christine Maswi}
\author[d]{Benard Chanzu}
\author[e]{Asaminew Teshome}
\author[e]{Bekele Kebebe}
\author[e]{Bekalu Tamene}
\author[e]{Samrawit Abebe}
\author[f]{Florian Pappenberger}
\author[f]{Matthew Chantry}
\author[g]{Isaac Obai}
\author[h]{Jesse Mason}

\affil[a]{Department of Physics, University of Oxford, Oxford, UK}
\affil[b]{National Centre for Atmospheric Science, Oxford, UK}
\affil[c]{IGAD Climate Prediction and Applications Centre, IGAD, Nairobi, Kenya}
\affil[d]{Kenya Meteorological Service Authority, Government of Kenya, Nairobi, Kenya}
\affil[e]{Ethiopian Meteorological Institute, Government of Ethiopia, Addis Ababa, Ethiopia}
\affil[f]{European Centre for Medium-Range Weather Forecasts, Reading, UK}
\affil[g]{Regional Bureau for Eastern Africa, United Nations World Food Programme, Nairobi, Kenya}
\affil[h]{United Nations World Food Programme, Rome, Italy}

\begin{document}

\twocolumn[ 
  \begin{@twocolumnfalse} 

\maketitle

\begin{abstract}
Ensemble forecasting has proven to be a vital tool for predicting extreme, life-threatening or only partially predictable weather events. As well as providing probabilistic products, individual ensemble members provide context and realisations of possible extreme weather. However, many National Meteorological Services in East Africa lack the computing resources required to run their local area models in ensemble mode over the full period of the two-week medium range. In this paper we test the performance of a short and medium-range forecast system recently put into daily use, comprising the global ECMWF ensemble forecast, post-processed using cGAN, a neural network model. We find that cGAN achieves probabilistic forecast skill comparable to that obtained using isotonic distributional regression (IDR), a recent statistical post-processing method, applied to either the ECMWF ensemble or the machine-learning models FuXi and GraphCast. Unlike IDR, cGAN generates individual ensemble-member realisations (potentially many thousands), including exceptional weather events. This ability was deemed important when assessed in an operational context by the National Meteorological and Hydrological Services of Kenya and Ethiopia. cGAN is
cheap to train/run and requires no additional post-processing. It is run on laptops making it practical as a high-resolution medium-range ensemble forecast system for Meteorological Services with limited computational resources.
\end{abstract}

\begin{adjustwidth}{35 pt}{35 pt}
\section*{Significance Statement}
We demonstrate useful gains in rainfall forecast skill by applying empirical corrections to physics-based forecast models, or to pure machine-learning models. Rainfall forecasts are required to predict flooding, storms and when to plant or harvest crops. This paper details the performance of models that have been developed for daily use at national meteorological centres and tested in operational forecast offices. The model tested here generates 1000 ensemble forecasts in a single forecast cycle on a standard desktop computer, not only providing more accurate predictions of the rainfall distribution than previously available, but also providing forecasters with individual realisations of possible severe weather events that may occur in the coming week.
\end{adjustwidth}

\vspace{0.35cm}
  \end{@twocolumnfalse} 
] 



\section*{Introduction}

Today in the East Africa region, the Ethiopian Meteorological Institute (EMI) and the Kenyan Meteorological Service Authority (KMSA) run deterministic rainfall forecasts every day out to 7 days using the NCAR Weather Research and Forecasting model (WRF) \cite{Skamarock2005,Skamarock2008} at 10km and 4km resolution respectively over local domains. NOAA's Global Forecast System (GFS), run by the United States' National Weather Service, provides initial and boundary conditions. For medium range prediction ICPAC (IGAD Climate Prediction and Applications Centre), based in Nairobi, Kenya, run an ensemble WRF twice weekly at 10km resolution, a similar setup to EMI but with initial and boundary conditions derived from the NCEP CFSv2 operational ensemble \cite{Saha2014}. When providing advice to users, these forecasts are complemented by external freely available resources from the UK Met Office, Météo-France and others. Shorter range nowcasting \cite{Roberts2022} information is also used via online products such as Forecasting African STorms Application (FASTA) \cite{FASTA} or Rain over Africa \cite{Hee2022,Amell2025}. Our aim is to complement these existing forecast products with large probabilistic rainfall ensembles that offer improved forecast skill and are readily accessible to operational forecasters.

The mission of National Meteorological and Hydrological Services (NMHS) is to provide accurate and timely information and guidance on upcoming weather risks. To inform operational rainfall forecasts that they issue over East Africa, a conditional Generative Adversarial Network (cGAN) \cite{Harris2022} has been developed to correct the ECMWF ensemble forecast (IFS) towards IMERG\footnote{Specifically the IMERG v7 {\it final} data set to represent the ``truth'' we are trying to forecast. See Materials and Methods.} blended satellite rainfall data. This code has been made robust, convenient and automated, and is in routine daily use by meteorologists in the region. Being computationally inexpensive, in a forecast cycle on a standard desktop computer, cGAN produces spatially correlated 1000-member ensembles. In this paper, we compare these ensembles to quantile mapping, which is commonly applied to correct distributions, isotonic distributional regression (IDR), which provides a strong recent statistical method, and to post-processed FuXi and GraphCast models, which provide a good machine-learning alternative to IFS. We find that both post-processing methods (cGAN and IDR) bring the physics-based IFS model and the machine-learning FuXi and GraphCast models to comparable levels of forecast skill. With skill extending to 7-day lead times, the resulting 6-hour and 24-hour accumulated rainfall forecasts are notably improved over high population areas in Kenya, Ethiopia, Uganda, Rwanda, Burundi and Tanzania, Lake Victoria and the Rift Valley lakes, mountainous regions, and the Indian Ocean. Biases are reduced to the climatological distribution in dry regions and over the Congo rainforest.

\subsection*{The region}

\begin{figure}
\centering
\includegraphics[width=\linewidth]{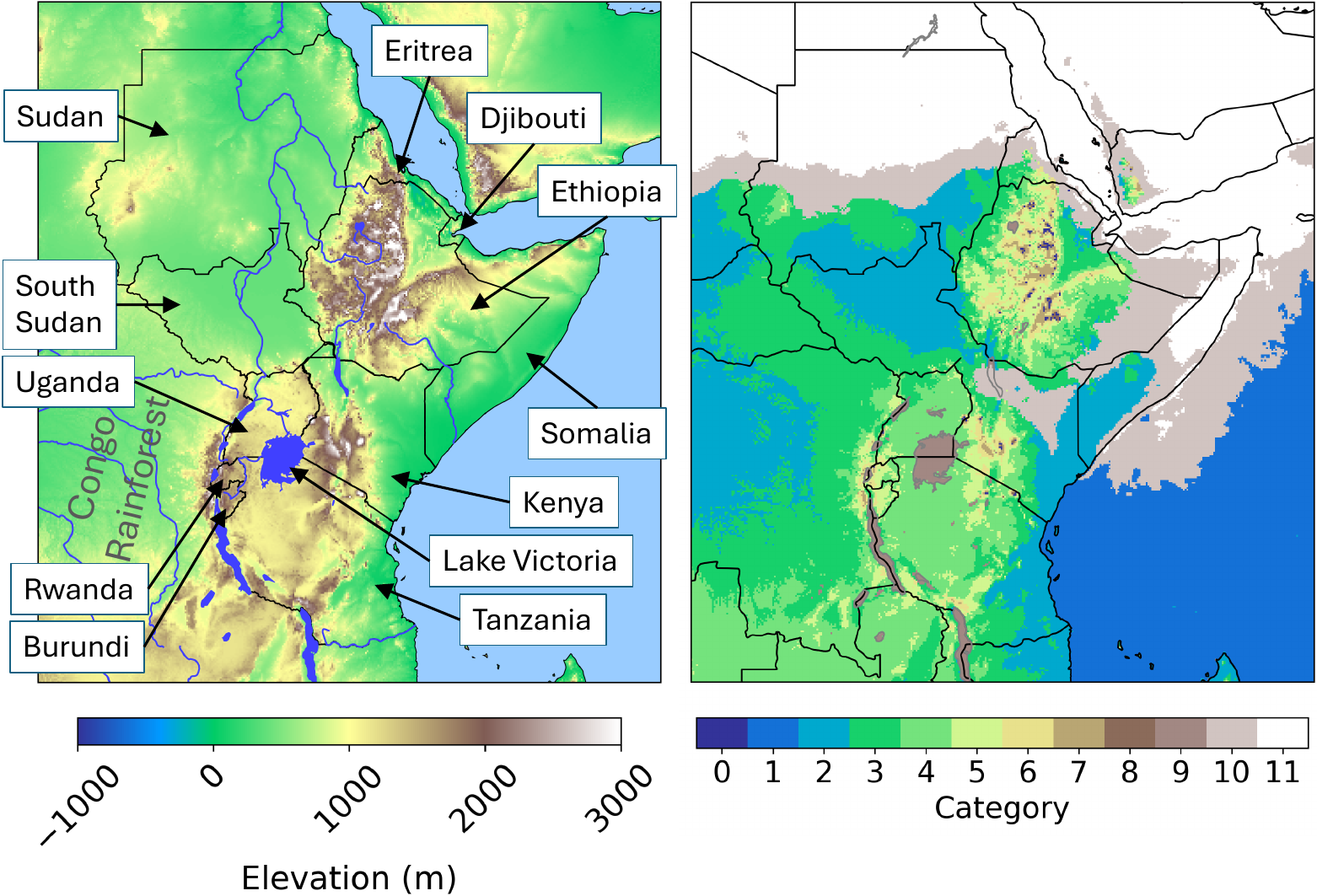}
\caption{{\bf Left:} The East Africa region. The forecast domain spans 13.7 degrees south to 24.7 degrees north and 19.1 to 54.3 degrees east. {\bf Right:} Grid points belonging to each category, chosen to broadly represent a particular physical regime, are combined to train a single IDR rainfall post-processing model, yielding a total of 12 models. See the Categories for IDR subsection in the Materials and Methods section for details.}
\label{fig:Elevation}
\end{figure}

This study focuses on tropical East Africa (Fig. \ref{fig:Elevation}). The Inter-Tropical-Convergence-Zone (ITCZ) appears as a band of rainfall in monthly-mean precipitation, migrating north and south across the region with the seasonal cycle and producing alternating wet and dry seasons \cite{Nicholson2017}. The topography has a large influence on the rainfall, bringing rain over the Ethiopian highlands and high regions of Kenya, Tanzania and the Congo rainforest. Regions exist to the north of Sudan and northern Kenya and Somalia are seasonally dry. Lake Victoria has a strong influence upon its surroundings, creating a local climate of heavy rainfall. See \cite{Palmer2023} for a review. Recent forecast studies have identified limited predictability of certain geographical rainfall structures in a single rainy season \cite{Kolstad2024}, while others have investigated the factors influencing the onset of the rainy season \cite{Gudoshava2022}. The time that rainfall typically occurs during the day varies across the region and it is useful for NMHS to have this information in their forecasts. As well as biases in the magnitude of rainfall, the IFS model has a bias towards convection too early in the day, which can be addressed to some extent by cGAN, \cite{Antonio2024}. Forecast skill assessments at medium \cite{Gudoshava2024} and monthly \cite{Endris2021} range have also been undertaken.

\subsection*{Physical modelling}

Weather models based on simulating the laws of physics, such as the Integrated Forecasting System, IFS \cite{IFS}, from the European Centre for Medium-range Weather Forecasts (ECMWF), produce predictions of the future state of the atmosphere. These predictions include random and systematic inaccuracies which stem from imperfect physical approximations within the model, for example the representation of convection \cite{Berthou2019}, and imperfect measurements of the model's initial conditions. To represent the uncertainty stemming from the random errors, ensembles of multiple model integrations are created by perturbing the initial conditions, time evolution and empirical physical parameters. Ensemble members that resemble a realistic physical situation are useful for practising forecasters in a couple of ways. Firstly, by looking at a selection of ensemble members the forecaster gains an intuition into the prevailing conditions leading to a particular probability. Will there be a sparse selection of rainfall events, with low probability at any particular location, or a low probability of one large event? Secondly, an ensemble bridges the difficult transition from thinking deterministically about the meaning of a single forecast, to thinking probabilistically about a forecast distribution.

\subsection*{``Pure'' machine-learning}

A different approach is to replace the physical simulation's dynamical evolution with a fully empirical model, for example the FuXi \cite{Chen2023} and GraphCast \cite{Lam2023} models in the deterministic setting, and FuXi-ENS \cite{Zhong2025} and GenCast \cite{Price2024} in the probabilistic ensemble setting. In all these cases the ERA5 reanalysis \cite{Hersbach2020} is the target truth, and ultimately this and the forecast initial conditions depend upon a physical forecast model. To see if the physical model is required at all, additional models are being developed based more directly upon measurements, for example GraphDOP \cite{Alexe2024}. ERA5 gives a poor approximation of IMERG rainfall measurements in the tropics and FuXi, GraphCast and GenCast perform poorly (see Fig. S4).

Other candidate pure machine-learning models include ECMWF’s AIFS Single \cite{Lang2024a}, operational on 25 February 2025, and AIFS ENS \cite{Lang2024b}, operational on 1 July 2025. AIFS ENS runs at 31 km resolution with known artefacts in accumulated precipitation. However, no full unseen test year  is available yet for either model within the periods considered here.

\subsection*{Post-processing}

Forecasts can be improved by identifying systematic errors relative to observations and developing empirical {\it post-processing} models to correct them \cite{Vannitsem2021,Bouallegue2023}. Post-processing 1-5 day tropical rainfall forecasts, especially in East Africa, is necessary to exceed the forecast skill of a climatological reference \cite{Vogel2020}. The distribution of measured rainfall has its peak at zero and decays with increasing rain. It is quite different from the distribution of rainfall forecast by IFS, with measurements typically having a lower chance of light rain or drizzle and fatter tails for a higher chance of heavy rain \cite{Lavers2021}. As an alternative to post-processing IFS with cGAN and to supply context, we also compare forecast skill when post-processing IFS with quantile mapping, and isotonic distribution regression.

\subsubsection*{Post-processing neural networks}

The cGAN takes into account conditional variables, including outputs from IFS other than rainfall. It produces spatially correlated ensemble realisations of possible rainfall fields, without attempting to approximate any correlation in time. Our code is adapted from the code originally documented in \cite{Harris2022} which was subsequently shown to improve forecast skill {of rainfall accumulated over 1h,} at 6h to 18h lead times over East Africa in \cite{Antonio2024}. The code has been improved and tailored to the 6h and 24h timescales out to 7 day lead times, and precise region documented here, see Materials and Methods. New models have been trained and the pipeline refined for efficient daily use by NMHS meteorologists. Here we expand on the diagnostic tests conducted in \cite{Antonio2024}, in particular assessing the quality of probabilistic forecasts using the CRPS.

More recently, diffusion-based post-processing models have been developed with the expectation that they are easier to train and can therefore converge to a more accurate solution. For example, nowcasting diffusion models are demonstrated in \cite{Leinonen2023,Mardani2025}. However, use of the Wasserstein loss function \cite{Arjovsky2017} has been found to mitigate training difficulties and the cGAN we are using has proven relatively easy to train. Another advantage of the cGAN is that it is very fast and enables production of 1000-member ensembles given the time and hardware available in East Africa. 1000-member ensembles are challenging with a diffusion model, though not impossible \cite{Li2024}.

\subsubsection*{Quantile mapping and IDR}

Quantile mapping, a common post-processing technique, in combination with multi-model forecasts has been found to extend rainfall prediction skill beyond climatology out to 9 days over Ethiopia \cite{Stellingwerf2021}. Quantile mapping corrects the forecast rainfall distribution of each forecast ensemble member to match that of the measurements \cite{Maraun2018}. For example, all forecasts of rainfall of around 4 mm/h, found to occur 0.1 percent of the time in the forecasts, might be mapped to around 6 mm/h measured to also occur 0.1 percent of the time.

One weakness of quantile mapping is that, in our example, a forecast rainfall of 4 mm/h does not always result in measurements of 6 mm/h. Instead, a distribution of rainfall is possible given the forecast. Recently, isotonic distributional regression (IDR), \cite{Henzi2021}, has been applied to estimate these distributions and to post-process ECMWF's IFS with some success in the region \cite{Ageet2023}.\footnote{If, as model output $r_{\text{model}}$ (being in our case the rainfall at a particular location) increases, the probability of the truth $r_{\text{truth}}$ exceeding some fixed threshold also increases or stays constant, for all choices of threshold, then the distribution $p(r_{\text{truth}} | r_{\text{model}})$ is said to be {\it isotonic}. IDR gets its power by assuming isotonicity. Although not all distributions are isotonic, it has been stated in \cite{Walz2024a}, that ``...estimators that enforce isotonicity tend to be superior to estimators that do not, even when the key assumption is violated...''.} IDR can be applied to estimate a discontinuous cumulative distribution function that, subject to the assumption of isotonicity, is optimal with respect to the Continuous Rank Probability Score (CRPS). It bypasses the production of an ensemble, producing the distribution directly, although it may also be applied to multiple ensemble members.

Using IDR to obtain a forecast rainfall distribution at each grid point, and nothing more, removes all spatial correlation information. Only the local 1D distribution is retained. This is not the disadvantage one might at first expect since many forecast users are only interested in the very local marginal distribution of rainfall and the collection of marginal distributions together retain spatial structure. Other uses, for example as the inputs to a hydrological model, might require catchment basin wide distributions including pixel-wise covariance information. For simple cases, this is still possible by application of IDR to correct the basin average model input.

\subsection*{Forecast evaluation}

Forecast performance is evaluated over a one-year test period against a probabilistic climatology \cite{Walz2024b}, as described in Materials and Methods. Results for an additional test year and further evaluation metrics are provided in the Supporting Information. Keeping these two years separate gives us some insight into how robust the results are. The quality of a forecast is assessed using statistics in the form of a {\it scoring rule}. ``A scoring rule is proper if the forecaster maximizes the expected score for an observation drawn from the distribution $F$ if he or she issues the probabilistic forecast $F$, rather than $G\neq F$'' \cite{Gneiting2007}. The Continuous Rank Probability Score (CRPS) is a popular and well established proper score and for that reason we apply it here. The CRPS quantifies the integrated squared difference between the forecast cumulative distribution and the cumulative distribution of an observation (a step function at the measured value). For a given ensemble forecast model, the expected CRPS reduces with more ensemble members \cite{Ferro2008,Leutbecher2019}. This is because the cumulative distribution, represented by the ensemble, becomes more precisely defined. The concept of a potential score that could be given with unlimited ensemble members, in conjunction with post-processing, then arises. In practice the IDR might be used to obtain a potential CRPS for model comparison \cite{Gneiting2025}. However, here we are interested in the skill of the forecasts that can be practically generated in time to issue an advisory, and so the standard CRPS is appropriate. Although the cGAN produces individual forecasts with a good spatial distribution \cite{Harris2022}, in this paper we focus on the quantitative assessment of the forecast distribution of local rainfall rather than the individual ensemble members. Spatial correlation is important in other contexts.

\section*{Results}

\begin{figure*}
\centering
\includegraphics[width=17.8cm]{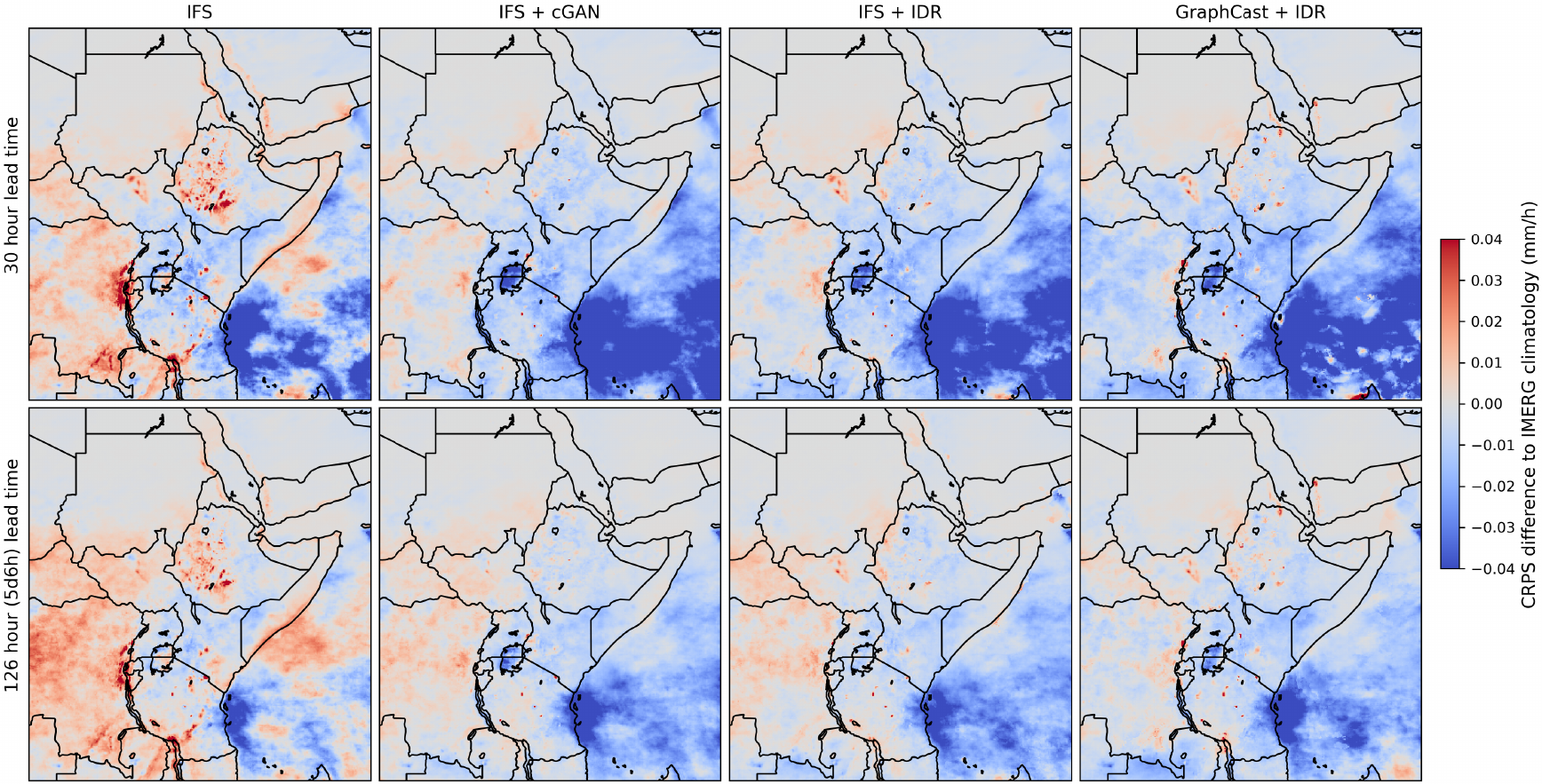}
\caption{Difference between the CRPS of the 24h rainfall accumulation forecasts and the CRPS of the IMERG climatological distribution at two example lead times, 30h (top) and 126h (bottom), averaged over one year (2023-6-1 to 2024-6-1) for {\bf left:} IFS, {\bf middle left:} IFS with cGAN post-processing to 1000 ensemble members, {\bf middle right:} IFS with IDR post-processing and {\bf right:} GraphCast with IDR post-processing. Blue means that the model has a lower (better) CRPS. Red means that the IMERG climatological forecast has a lower CRPS. For robustness, compare to Fig. S3.}
\label{fig:CRPS_maps}
\end{figure*}

Differences between the CRPS of the IMERG v7 {\it final} probabilistic climatological benchmark (see Materials and Methods) and the one-year mean CRPS of each forecast, indicate where the forecast outperforms climatology (also see Materials and Methods for training and test periods). Maps of these differences for 24h rainfall accumulation periods (06:00 to 06:00 UTC) are plotted in Fig. \ref{fig:CRPS_maps}. At 30 hours lead time after forecast initialisation the IFS forecast has a better (lower, blue) CRPS than climatology in parts of Kenya, Uganda and Tanzania, as well as over the southern portion of the Indian Ocean within this domain. However, over large areas the IFS forecast CRPS is higher (worse, red). These include the entire vicinity of the Congo rainforest, mountainous regions (compare to Fig. \ref{fig:Elevation}), and northern Somalia. At 5 days lead time, predictability over climatology has reduced. There are large red areas along the coast of Somalia and the westernmost countries in the region, although CRPS errors over mountains have reduced.

Applying cGAN post-processing to the IFS forecasts improves the CRPS across much of the study region. Areas to the east of the Congo rainforest and south of Sudan show a lower CRPS than IMERG climatology. Rainfall over the mountainous regions of Ethiopia\footnote{However, there is evidence that the result in western Ethiopia is not robust; see Fig. S3.}, Tanzania and surrounding Lake Victoria has been corrected, as has the rainfall over Lake Victoria itself. The forecast now has skill over the entire Indian Ocean residing within the domain. The CRPS over the Congo rainforest has been improved, although it is still not robustly better than the IMERG climatological forecast. Little or no improvement is shown in northern Somalia, southern Sudan and dry desert areas to the north. See Fig. S5 for the 6h accumulation equivalent.

\begin{figure}
\centering
\includegraphics[width=\linewidth]{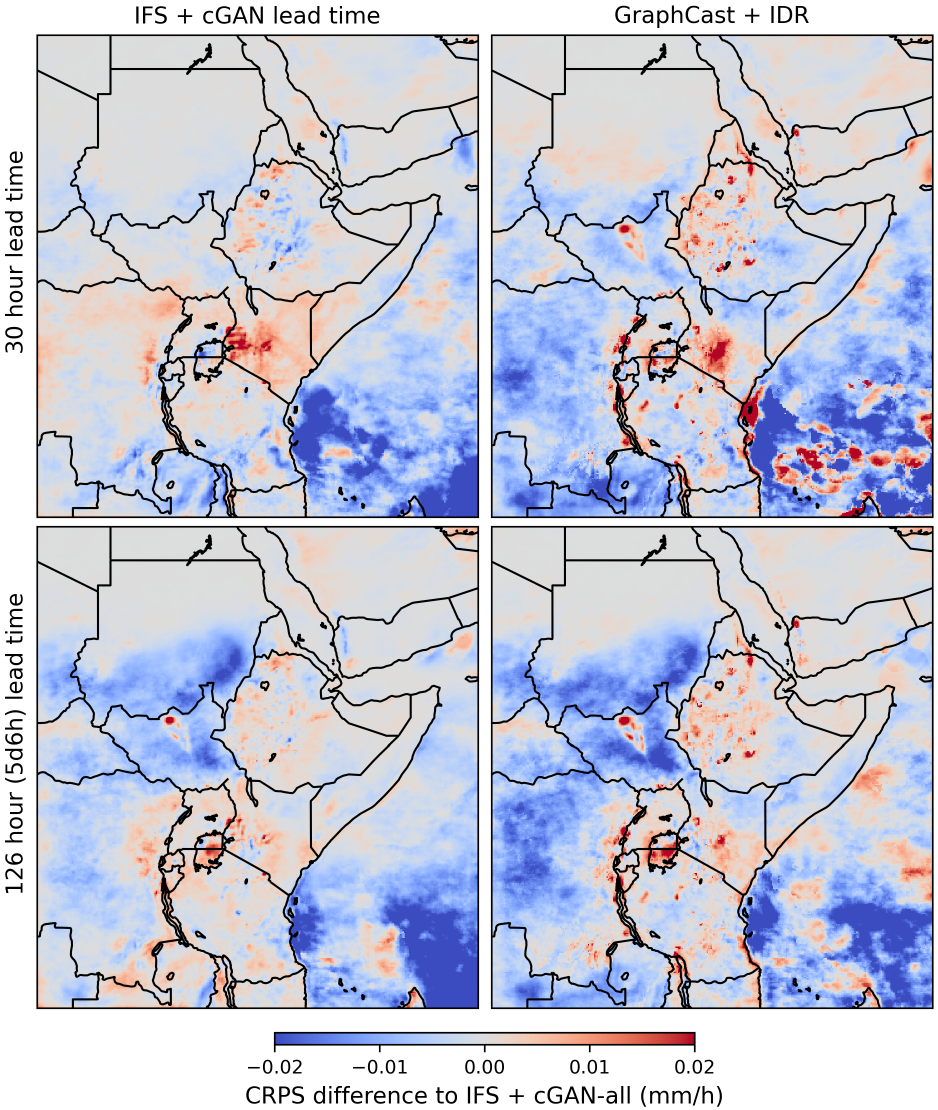}
\caption{Difference between the one-year mean CRPS of the 24h rainfall accumulation of 1000 ensemble members of the IFS+cGAN forecast trained on data from all forecast lead times, denoted {\it IFS+cGAN-all} and {\bf left column:} IFS+cGAN trained on data only from the lead time plotted, denoted {\it IFS+cGAN lead time} and {\bf right column:} single ensemble member (deterministic) GraphCast with IDR post-processing to obtain a distribution. Blue means that the labelled model has a lower (better) CRPS. Red means that the {\it IFS+cGAN-all} forecast has a lower CRPS. For robustness, compare to Fig. S7.}
\label{fig:cGAN1000}
\end{figure}

IDR shows similar improvements to the time mean CRPS, very comparable in magnitude and pattern to those made by cGAN. However, differences in CRPS between IFS+cGAN and IFS+IDR at individual forecast lead times suggest that the two methods correct different aspects of the rainfall distribution. Very similar improvements to the CRPS again are produced by using the GraphCast model with IDR post-processing. The CRPS in the region of the Congo rainforest is improved with respect to the other methods. GraphCast performs less well over mountainous regions, potentially because its relatively coarse native resolution ($0.25^{\text{o}}$) does not adequately resolve convective processes over complex terrain \cite{Mwanthi2024}. Maps emphasising the differences between IFS+cGAN and GraphCast+IDR are given in Fig. \ref{fig:cGAN1000}. The broad patterns partly reflect that, absent significant topographical features, the local rainfall probability distribution changes slowly in space. However, it is difficult to distinguish this from a general slight wet or dry bias without further work, see Fig. S1.

We suspect that the large region of poor CRPS in South Sudan, and the isolated regions of poor CRPS (red dots) in Tanzania, which appear over regions of swamp, present in many forecasts are due to problems with IMERG. These patterns are also present in the annual mean IMERG rainfall where such local increases are not reported, see Fig. S1. There are also red dots of poor CRPS over Mount Kenya and Mount Kilimanjaro where rainfall is locally higher. An effective post-processing correction towards IMERG might be expected to remove these apparent errors, when assessed against IMERG itself, even when IMERG is not a good representation of reality. To some extent, in the Sudd region of South Sudan, this appears to have happened, see Figs. \ref{fig:CRPS_maps} and S3. However, cGAN is designed to be spatially invariant, utilising data from other regions to correct a given location, see Materials and Methods. IDR is applied uniformly over spatial categories, Fig. \ref{fig:Elevation} and table \ref{table:dataCategories}. Both methods appear to either not have access to the data required, or not be sufficiently detailed, to model these special cases.

\subsection*{Domain average CRPS and model differences}

\begin{figure}
\centering
\includegraphics[width=\linewidth]{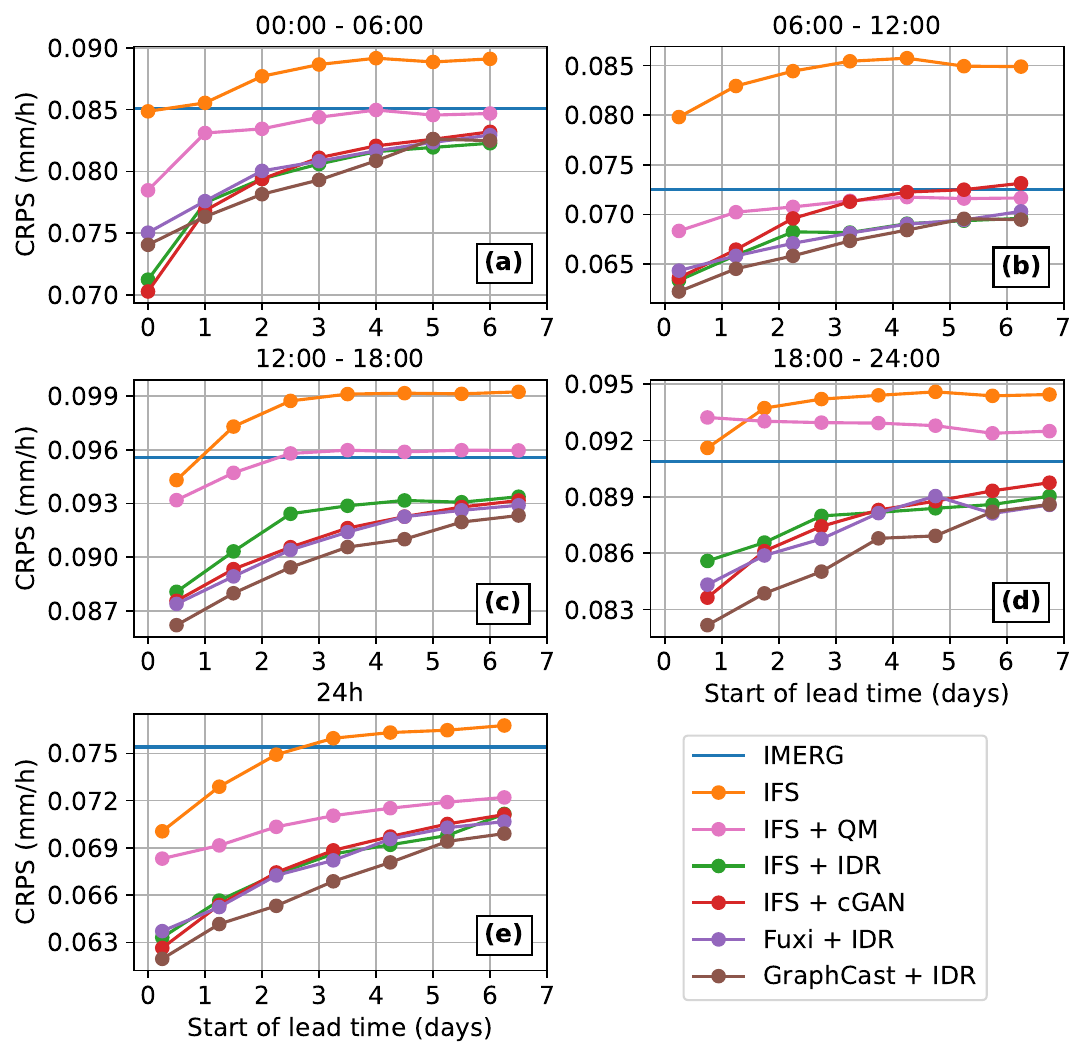}
\caption{6h and 24h rainfall accumulation CRPS values averaged over the East African domain and the one-year test period (2023-6-1 to 2024-6-1). The top four plots (a,b,c,d) represent the four different 6-hour rainfall accumulation periods per day. Times are UTC and all forecasts are initialised at 00:00 UTC. The bottom plot (e) represents rainfall accumulation from 06:00 to 06:00. Each point represents the start of the accumulation period. QM stands for quantile mapping applied to each of the 50 ensemble members. 1000 ensemble members are generated for the IFS+cGAN points. The blue line is the CRPS of the IMERG climatological distribution. Lower is better, but the domain average masks important issues, see Fig. \ref{fig:cGAN1000}.For robustness, compare to Fig. S6. See also Figs. S14 and S15 for equivalent plots restricted to each country.}
\label{fig:CRPS_mean}
\end{figure}

To summarise the spatial CRPS patterns shown in Fig. \ref{fig:CRPS_maps}, we calculate domain-averaged values. Although this removes spatial detail, it allows the evolution of CRPS with forecast lead time to be compared across models in a single plot. Plots for each of the four 6h accumulation periods in a day and the 24h accumulation period are given in Fig. \ref{fig:CRPS_mean}. With this metric IFS forecasts become worse than climatology after less than 1 day for 6h accumulated rainfall (fig. \ref{fig:CRPS_mean}a,b,c,d) and after 2 to 3 days for 24h rainfall accumulations (fig. \ref{fig:CRPS_mean}e). Quantile mapping (QM) improves on IFS alone, bringing the mean CRPS closer to or below climatology.

The remaining models (IFS+IDR, IFS+cGAN, FuXi+IDR, GraphCast+IDR) are below (better than) climatology even for 6h accumulated rainfall, and if the trend continues skill would extend beyond seven days. The models achieve similar mean CRPS values, with two notable exceptions: for the 6h accumulations from 06:00 to 12:00 (fig. \ref{fig:CRPS_mean}b), cGAN post-processing appears to underperform, and for the 6h 12:00 to 18:00 (fig. \ref{fig:CRPS_mean}c), IDR post-processing seems not to do so well. GraphCast+IDR slightly outperforms the other models with this metric. Figure \ref{fig:cGAN1000} reveals that this is due to performing well over the western part of the domain, the Congo rainforest, Sudan and South Sudan, and perhaps the Indian Ocean. Unfortunately, this is not the case in the highly populated regions of Kenya, Ethiopia, Eritrea and Uganda where IFS+cGAN outperforms GraphCast+IDR.

To illustrate the differences between models more clearly, in Fig. \ref{fig:cGAN1000} we change the baseline from the IMERG climatology to the forecast provided by IFS+cGAN where cGAN is trained on all lead times, as opposed to the version used in Figs. \ref{fig:CRPS_maps} and \ref{fig:CRPS_mean} which consisted of individual models each trained to specialise on a single lead time, see Materials and Methods. Forecasts over the Indian Ocean appear to be improved by training at each lead time, while forecasts over Kenya are improved by combining all lead times into model training. Inconsistent results for GraphCast+IDR over the Indian Ocean are quite large and appear to contribute overall to the area mean CRPS illustrated in Fig. \ref{fig:CRPS_mean}. Comparison to Fig. S7 shows that the Indian Ocean differences between models is not robust. This suggests that the difference in the area mean CRPS is somewhat due to individual weather systems. Country wide averages of the scores are presented the Supporting Information in Figs. S14, S15, S17 and S18.

\subsection*{Forecast distribution}

\begin{figure}
\centering
\includegraphics[width=\linewidth]{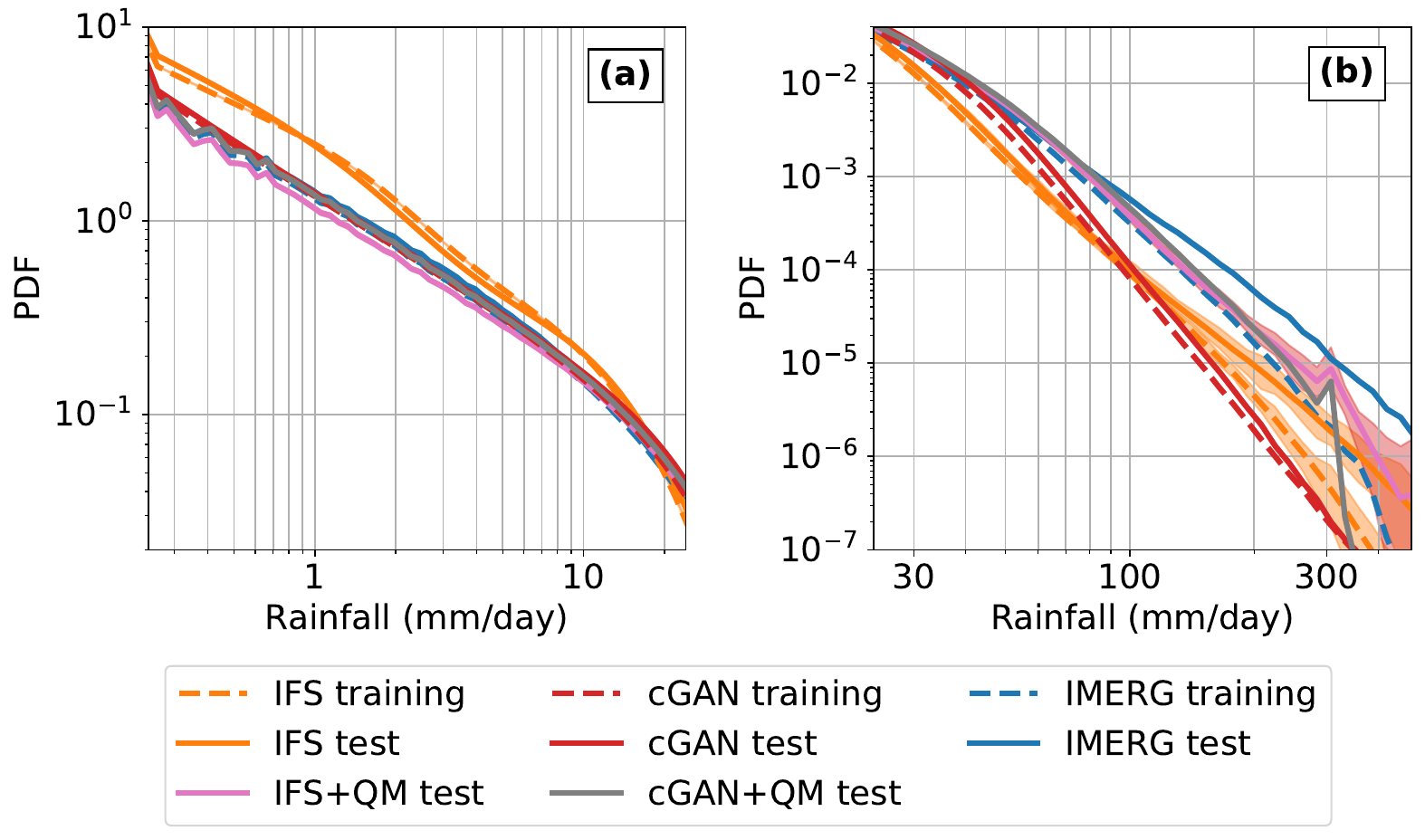}
\caption{Histograms of the rainfall in the test period (2023-6-1 to 2024-6-1) averaged over 24 hours using a 30h to 54h lead time. {\bf (a) Left:} Rainfall below 1 mm/h. {\bf (b) Right:} Rainfall above 1 mm/h. The dashed lines indicate the distribution over the model training period (2018-1-1 to 2022-1-1). The solid lines indicate the distribution over the model test period. Note the different logarithmic axes. The lines labelled cGAN indicate cGAN applied to post-process the IFS and the grey line indicates a further correction by quantile mapping, IFS+cGAN+QM.}
\label{fig:QM}
\end{figure}

The local distribution of rainfall, accumulated over 24h between 30h and 54h lead times, is averaged across the whole domain and over both the four-year training and one-year test periods (see Materials and Methods). This is plotted in Fig. \ref{fig:QM}. Although they are imperfect representations of the true rainfall distribution, see Materials and Methods, the IMERG training and test distributions are similar (blue lines). The logarithmic scale in the right-hand plot (b) exaggerates the tiny differences in the tails. We assume that the IMERG climatology has not changed much between the test and training data, and that these two curves illustrate the size of the difference caused by random or systematic sampling error. That is, different numbers and intensities of weather events in the training and test data sets and changes over time to the characteristics of the satellites and their instruments.

The IFS training and test distributions are also similar (orange lines), although, except for the very heaviest rainfall, they clearly differ from the target IMERG distribution. These differences might be narrowed by resolving convection explicitly \cite{Berthou2019}. The range of values, most visible in the right-hand plot (b), indicates the maximum and minimum probability density over all ensemble members, which is smaller than the random sampling error, suggesting overconfidence of the IFS. The IFS training data with quantile mapping applied (not shown) accurately follows the IMERG training distribution in this plot. The IFS test data with quantile mapping applied (pink line) closely follows the IMERG distribution: in the tails it is within the IMERG sampling error. cGAN applied to IFS (red lines) corrects the low rainfall part of the distribution well. However, as noted in \cite{Antonio2024} for 1h accumulations, cGAN underestimates the probability of rainfall in the high rainfall part of the distribution. Above $\sim 80$ mm/day, cGAN underestimates rainfall more than IFS. The uncertainty range on the cGAN ensemble members (not shown) is similar to the range of the IFS ensemble members. As in \cite{Antonio2024}, applying quantile mapping to the cGAN output (brown line) corrects the tail of the distribution, although we notice in this work that this comes at the cost of an increased CRPS (see Fig. S9). The jagged artifacts near zero rainfall present in IMERG do not appear in cGAN.

\section*{Discussion}

We have evaluated an AI-based post-processing system that improves global rainfall forecasts over East Africa and has been used routinely by local NMHS over the past three years. Building on earlier work \cite{Harris2022,Antonio2024}, here the cGAN has been substantially upgraded, extended to seven day forecast lead times and verified against a probabilistic climatology. Trained especially for rainfall over East Africa, it corrects high resolution physical model outputs that could be important to resolve the deep convective systems driving rainfall over the region, whereas most other products are trained and run on coarser 25 km grids using reanalysis datasets that need additional correction. Because it runs on a standard desktop computer without requiring a GPU, the system represents a substantial improvement in forecasting capabilities for regions with limited computational infrastructure. We show its overall improvements and compare to three very different methods of rainfall forecasting, which have been shown (Fig. \ref{fig:CRPS_maps}) to achieve approximately the same improvement in CRPS, with IFS+cGAN-all being slightly better over populated regions (Fig. \ref{fig:cGAN1000}).

Although further improvements remain possible, the relatively small differences in CRPS among the post-processed models suggest diminishing returns from additional empirical refinements. We therefore expect that further gains in forecast skill may be incremental, whereas improvements comparable in size to those achieved by applying cGAN to IFS may require new methodological insights. Other findings include the fact that separate cGAN models trained individually on each lead time can sometimes outperform more general cGAN models trained to predict all lead times, and that cGAN underestimates infrequent heavy rainfall. Correcting for this exposes a potential trade-off between the forecast reliability and its probabilistic resolution.

The conditional Generative Adversarial Network (cGAN) is competitive with isotonic distributional regression (IDR) and has some advantages. One potential advantage is that the generated ensemble members retain spatial correlations, which may allow their application to different use cases without requiring additional model training. However, this potential benefit has not been evaluated here. Another advantage is the conventional ensemble output, which more gently bridges the conceptual gap between deterministic and probabilistic forecasts. However, this is not useful to all users and IDR can also be adapted to cover many use cases. Another feature of cGAN is its low computational cost: 1000 ensemble members can be produced with 7 lead times in less than 30 minutes on a standard desktop computer. Unlike full atmospheric machine-learning models such as GraphCast, cGAN does not require a GPU. This makes cGAN practical for producing multiple probabilistic forecasts per day using hardware readily available in East Africa.

Another consideration is the low computational cost of model training. Both IDR and cGAN can be trained using modest hardware. If steps are taken to limit data usage, IDR can be practically trained in a short period using a CPU. A single A100 GPU was used to train all the cGAN models used here, taking between around one and three days per model. This enables us to innovate and experiment with different lead times, input variables and network sizes at a low cost. For example, when the CAPE variable from IFS was replaced with the more physically consistent MUCAPE, that was not in our training data, we were able to maintain the same CRPS skill by retraining the model using IMERG climatology as an alternative input.

Figure \ref{fig:cGAN_progress} indicates that for the area mean CRPS, models trained on each lead time separately, more often than not, perform better. However, Fig. \ref{fig:cGAN1000} suggests firstly that optimising the overall area mean score might not provide the most useful model, and secondly there may be a trade-off in that different models are best applied only to the specialist regions that they are good at. In our case if one is interested in 24h accumulated rainfall over Kenya and Uganda, the IFS+cGAN-all model, trained on all lead times, achieves the best CRPS of the models considered here. Figures \ref{fig:CRPS_maps} and \ref{fig:CRPS_mean} indicate that differences in forecast skill among the models are relatively small, and at the colour scale used in Fig. \ref{fig:cGAN1000} the impact of a small number of random individual weather events becomes important. Compare to Fig. S7 and to seasonal breakdowns, Figs. S10, S11, S12 and S13.

Additional data would allow a more robust assessment of these results. In particular we plan to make use of the 20-year hindcasts of the latest operational IFS retained at ECMWF. Additional years of test data, beyond the two test periods considered here would allow more detailed analysis regarding locally and seasonally varying skill, and to quantify some of the uncertainties in our analysis. We would also expect that cGAN or IDR trained on the current IFS version would lead to some improvements in skill. One might think that more training data will lead to better IDR and cGAN models. However, it is not clear that this is true. To save on computational costs, IDR was severely limited to a small subset of the available data already and we have noticed that when training cGAN, skill seems to stop improving before our current 4-year training data set is covered.

Further improvements in forecast skill may be achieved through higher-quality observational datasets. IMERG v7 is a high quality product for our purposes. It is easy to use, has the appropriate spatial and temporal resolution, and is based on a wealth of measurements. However, it still has biases, with some data suffering high uncertainties. Additional rain gauges not used within IMERG exist. Incorporating them with improved quality control, and further calibration of IMERG \cite{Funk2022} or similar products \cite{Papacharalampous2023}, improvements can be achieved. Improvement of post-processing models might also be achieved by restricting to only training where measurements are high enough quality. Further down the line, ground-based radars are coming online in the region.

It is possible to decompose the CRPS into reliability and resolution \cite{Candille2005} and more recently additional decompositions have been proposed \cite{Arnold2024}. For the models considered here, it would be interesting to find out in more detail the relative contributions to good forecast performance. The loss function used to train, or optimise, the cGAN, which is constructed from the Wasserstein metric \cite{Gulrajani2017}, and the CRPS optimised by IDR, are both known to not place emphasis on the tails of the distribution. As a result, we see in histograms of the distribution under-representation of the tails, Fig. \ref{fig:QM}. This is because a relatively small amount of probability is represented in the part of the distribution corresponding to these rare events. It might even be the case that physics-based models are currently better at simulating the most extreme record-breaking events than machine-learning approaches \cite{Zhang2025}. Our ability to correct the distribution with quantile mapping indicates a potential trade-off between reliability and probabilistic resolution of our forecasts. Optimising for the mean distribution (with quantile mapping) does not optimise for the CRPS. The approach of evaluating a model on a select subset of extreme events with limited data has the downside of discrediting skilful forecasts \cite{Lerch2017}. As mentioned in \cite{Lerch2017}, a way forward might be to use weighted scoring rules that emphasise the tails. Another approach is to specialise a model on a single probability of exceeding some user-defined, potentially extreme, threshold \cite{Nath2026}.

Local area forecast models have been specialised on particular locations for many years. In an analogous way, given the relatively low computational costs, it appears that we are approaching the point where data-driven forecast post-processing models can be individually optimised to address the different statistical needs of individual users, as also discussed in the context of user-centred forecasting systems \cite{Alfieri2012, Demeritt2013}.

\footnotesize
\section*{Materials and Methods}

Our training data consists of 50 members of the IFS operational ensemble forecast, linearly interpolated from the native IFS grid \footnote{The IFS medium-range ensemble ran at $\sim$ 18 km resolution until 27 June 2023, when IFS Cycle 48r1 increased the resolution to $\sim$ 9 km; this falls 26 days into our first test period. Cycle 49r1, implemented on 12 November 2024, activated the stochastically perturbed parametrisations scheme, altering ensemble spread, and falls within our second test period. These changes work against the post-processing rather than for it: the reported gains are therefore plausibly a lower bound on what retraining against the current cycle would achieve.} to the $\sim 11$km longitude/latitude grid specified by the IMERG v7 {\it final} observational data, see below. We consider 24h and 6h rainfall accumulation periods, chosen to match those set by the NMHS to align with the 06:00 UTC WMO standard for rain gauge measurements \cite{WMO2018}. Data is restricted to our region (Fig. \ref{fig:Elevation}) and for model training and validation we use forecasts initialised daily at 00:00 UTC over a period of 4 years, 2018 to 2021 inclusive. To test the post-processing models, we use the same data but between the 1st of June 2023 and the 1st of June 2024. Although leaving a gap of 1 year and 5 months between training and test data sets is prudent, the main motivation was to make practical use of data already downloaded and to have a test data set that does not overlap with the training data of other machine-learning models. To gain some insight into the robustness of the results, analysis of an additional year of test data, between the 23rd of September 2024 and the 23rd of September 2025, is included in the Supporting Information for comparison. The models reported here were trained before data from this additional test year became available.

For evaluation of the FuXi and GraphCast models, we use forecasts generated by operational versions available from ECMWF. These were trained on ERA5 at a 0.25$^\text{o}$ resolution and fine-tuned on the ECMWF HRES forecast. A summary of all models compared is provided in table S1 in the supporting information.

\subsection*{IMERG data and climatological benchmark}

We use the IMERG version 7 final data set \cite{Huffman2020,Huffman2023} to represent the ``truth''. The IMERG data is derived from a combination of sources to provide 30-minute rainfall accumulations in 0.1-degree grid boxes over the entire tropical region. We take a subset over East Africa, see Fig. \ref{fig:Elevation}. Recent reviews of the performance of IMERG \cite{Ageet2022,Rajani2022,Xiong2025} indicate that it retains substantial uncertainty. It struggles with drizzle and moderate-to-extreme events, particularly those exceeding 5 mm/h, and we noticed issues over swamp regions and South Sudan. Earlier versions of IMERG have been used in East African regional forecast studies before \cite{Antonio2024,Ageet2023} and although it was not always the leading product, it has stood up relatively well to comparison with regional rain gauge data \cite{Omay2025}.

To assess how good the cGAN forecast is we require a difficult to beat climatological probabilistic benchmark forecast. Following \cite{Walz2024b}, an IMERG record (accumulated for 6 or 24 hours depending upon the forecast assessed) is chosen from the forecast time of the year up to $\pm$15 whole days from a year between 2001 to 2021 inclusive. This constitutes one climatological ensemble member. The process is repeated for all available times to build up a climatological ensemble forecast with 651 members. The advantage of this procedure is that we build up a good climatological distribution. Note that we are comparing the practical forecast skill obtained and not the {\it potential} forecast skill that would be obtained with an infinite ensemble. Therefore, having 651 ensemble members in the climatological forecast is not an unfair comparison to a 50-member IFS forecast. A map of the CRPS of the IMERG benchmark during the test period is given in Fig. S1.

\subsection*{Categories for IDR}

Dry conditions prevail during much of the four-year training period. When rainfall events do occur they are correlated in time. For each 11km grid box we therefore have a very limited number of independent events with which to train quantile mapping or IDR post-processing models. However, the study region is large and rainfall forecasts at different locations may exhibit similar biases and uncertainties. We therefore aggregate training data from across the region using two strategies. The first is to select points from the entire region, with each grid box at each time having equal probability of being selected. The second divides the region into categories. Training data is then selected exclusively from a single category and a model specialist to that category is trained. The criterion is firstly our assumption that each category represents a particular physical situation. Model variables within a category are expected to share biases to some extent. Secondly, categories were chosen to contain a reasonable quantity of non-zero rainfall measurements. Drier regions have more grid points to compensate somewhat for the infrequent rainfall. The categories chosen here are based on elevation and are listed in Table \ref{table:dataCategories}, and although not optimal, were found to improve the CRPS of the IDR. The same problem is addressed in \cite{Antonio2024} for application to quantile mapping by dividing the data into square regions of longitude and latitude.

\begin{table}[t!]
\begin{tabular}{p{0.11\linewidth}p{0.64\linewidth}p{0.11\linewidth}}
Category & Description & No. grid boxes \\
\midrule
All & All grid boxes selected with equal probability & 135168 \\
\midrule
0 & Elevation is below sea level (0m) & 121 \\
1 & Ocean & 20242 \\
2 & Elevation between 0m and 500m & 17368 \\
3 & Elevation between 500m and 1000m & 21363 \\
4 & Elevation between 1000m and 1500m & 15284 \\
5 & Elevation between 1500m and 2000m & 3992 \\
6 & Elevation between 2000m and 2500m & 1448 \\
7 & Elevation between 2500m and 3000m & 557 \\
8 & Elevation greater than 3000m & 38 \\
9 & Lakes substantially above sea level: Land-sea mask greater than 50\% and elevation greater than 100m & 1367 \\
10 & Low rainfall regions: IMERG training data rainfall average between 0.025 and 0.05 mm/h & 15058 \\
11 & Very low rainfall regions: IMERG training data rainfall average below 0.025 mm/h & 38330 \\
\bottomrule
\end{tabular}
\caption{{\bf Rainfall model regions.} Data from the region is split into the categories listed above. Locations that satisfy the criteria for multiple categories are only included in one, with the priority being categories nearest the bottom of the table. The low rainfall regions vary depending upon the time of day and season. Example number of grid boxes for the full 4-year averages are given and an example map indicating the categories are shown in Fig. \ref{fig:Elevation}, which bears some resemblance to the K\"oppen-Geiger climate classification \cite{Beck2018}.}
\label{table:dataCategories}
\end{table}

\subsection*{Quantile mapping and IDR post-processing}

To train the quantile mapping, in each case (category and lead time) $10^4$ data points were used for consistency with IDR detailed below. Each IFS or cGAN ensemble member was mapped to reproduce the IMERG rainfall distribution. As can be seen in the right panel of Fig. \ref{fig:QM}, the quality of the mapping then seems to be dominated by the limited number of physical weather events in the extremes. We found negligible differences to the area mean CRPS between the quantile mapped model built using training data from the whole region, and the 12 models defined using regional categories above (presented here). Each quantile mapping model is specialised to a specific lead time except for 24h accumulations, where the quantile mapping model defined for 126-hour (5 days and 6 hours) lead times is also used for 150-hour (6d6h) lead times.

IDR as described in \cite{Walz2024a} and applied here does not require the optimisation of any tuning parameters. However, in practice there are choices to be made with its application. Firstly, there is a trade-off between the quantity of data used to estimate the distribution and how specific a situation that data applies to. For example, in the case considered in this paper, an independent IDR model at each grid point required an impractical level of storage and did not yield good results. This problem can be overcome by combining data into the categories defined above and fitting a single model to each one. A combination of grid points, randomly chosen from one of the categories shown in Fig. \ref{fig:Elevation}, increases the quantity of training data, and reduces how specific IDR models can be. However, the limitation then becomes computational time and memory usage which increases super-linearly with the number of data points. On our computers, the practical limit for the number of training data points for fitting a deterministic model was around $10^4$. We therefore selected $10^4$ training data points by randomly selecting within a category, a longitude, a latitude and a time for each point. We found that splitting the data into categories listed in Table \ref{table:dataCategories}, marginally improved our results over simply using the entire region. As for quantile mapping, each IDR model is specialised to a specific lead time except for 24h accumulations, where the IDR model defined for 126-hour (5d6h) lead times is also used for 150-hour (6d6h) lead times.

The IDR algorithm is applicable to ensemble distributions. However, this comes at a high computational cost. We fit an IDR model to the IFS ensemble mean rainfall, another model to the joint ensemble mean and ensemble standard deviation of rainfall, and another to 10 ensemble members. Given our forecast time limitations, we found fitting to 50 ensemble members (with $10^4$ fifty element vectors) computationally impractical. We expected that the IDR corrected distribution would clearly depend upon the ensemble spread. However, these three models had very similar performance. The ensemble mean only model is much faster to fit and provides all our plotted results.

\subsection*{The conditional Generative Adversarial Network (cGAN)}

Building on the cGAN code developed in \cite{Harris2022} and applied to East Africa in \cite{Antonio2024}, several refinements have been made. As training data, multiple physical variables are taken from the ECMWF IFS model ensemble output, see Table \ref{table:inputs}. These variables, which differ from those used in \cite{Harris2022} and \cite{Antonio2024}, were selected based upon expert judgement and trial and error, with some additions from tests using linear regression. We use regional images of the IFS ensemble mean and standard deviation of each variable. This represents the ensemble distribution more accurately and was indeed found to improve the CRPS relative to the HRES deterministic forecast fields used in \cite{Harris2022} and \cite{Antonio2024}. $128\times128$ grid point training samples are selected randomly from the full $384\times352$ set of grid points from each date. The aim is to maintain some spatial invariance and to increase the number of training samples. In \cite{Harris2022} the IFS forecast was downscaled to ($\sim$1km) ground-based radar data representing the truth. In \cite{Antonio2024} the ground-based radar, which is not available in East Africa, was swapped for the IMERG v6 product. Here we switch to the improved IMERG v7 final product and switch back to the logarithmic representation of precipitation used in \cite{Harris2022}. The forecast period was extended from the hourly accumulations of rainfall over lead times of 6-18 hours used in \cite{Antonio2024}, to 6h and 24h accumulation periods out to seven days, which are favoured by regional meteorologists. Separate models were then trained for each lead time and for a combination of all lead times. In forecast mode, each cGAN model outputs an image of a random rainfall field, drawn from the (approximate) distribution of possible forecasts. This image constitutes a single ensemble member, and the process is repeated to build a large ensemble forecast.

\begin{table}[t!]
\begin{tabular}{lp{0.65\linewidth}}
ECMWF code & Predictor \\
\midrule
CAPE & Convective available potential energy (6h model trained using all lead times only$^1$) \\
CP & Convective precipitation \\
TP & Total precipitation. (Used as the IFS prediction of rainfall.) \\
MCC & Medium cloud cover. (Cloud cover at medium altitude.) \\
SP & Surface pressure \\
SSR & Surface incoming solar radiation \\
T2M & Two metre temperature \\
TCIW & Total column ice water. (The total ice water in a column of air within the grid box and between the surface and the top of the atmosphere.) \\
TCLW & Total column liquid water \\
TCRW & Total column rain water \\
TCW & Total column water \\
TCWV & Total column water vapour \\
U (700 hPa) & Zonal (West to East) wind at 700 hPa \\
V (700 hPa) & Meridional (South to North) wind at 700 hPa \\
Elevation & Nearest neighbour interpolated to the IMERG grid from the 30-arcsecond Global multi-resolution terrain elevation data 2010 (GMTED2010) \cite{GMTED2010}.\\
Land-sea-mask & Average fraction of water, including ocean lakes and rivers, within an IMERG grid cell. Sourced from $\sim$10 m resolution ESA world cover 2020 data \cite{ESAWorldCover2020} \\
IMERG climatology & The climatological mean and variance derived from the IMERG distribution defined for the climatological benchmark. (6h models trained separately on each lead time only$^1$) \\
\bottomrule
\end{tabular}
\caption{{\bf cGAN inputs} Inputs to the cGAN model sourced from the IFS forecast model outputs unless otherwise indicated. The ECMWF parameter codes can be found at \url{https://codes.ecmwf.int/grib/param-db}. \\
$^1$ CAPE was removed from the operational IFS model outputs after the 6th of June 2024, therefore we removed CAPE from cGAN models used for operational post-processing. The 6h cGAN model trained with CAPE and using all lead times is retained here for comparison. This resulted in reduced CRPS skill which was approximately regained by supplying the IMERG climatology instead.}
\label{table:inputs}
\end{table}

\begin{figure}
\centering
\includegraphics[width=\linewidth]{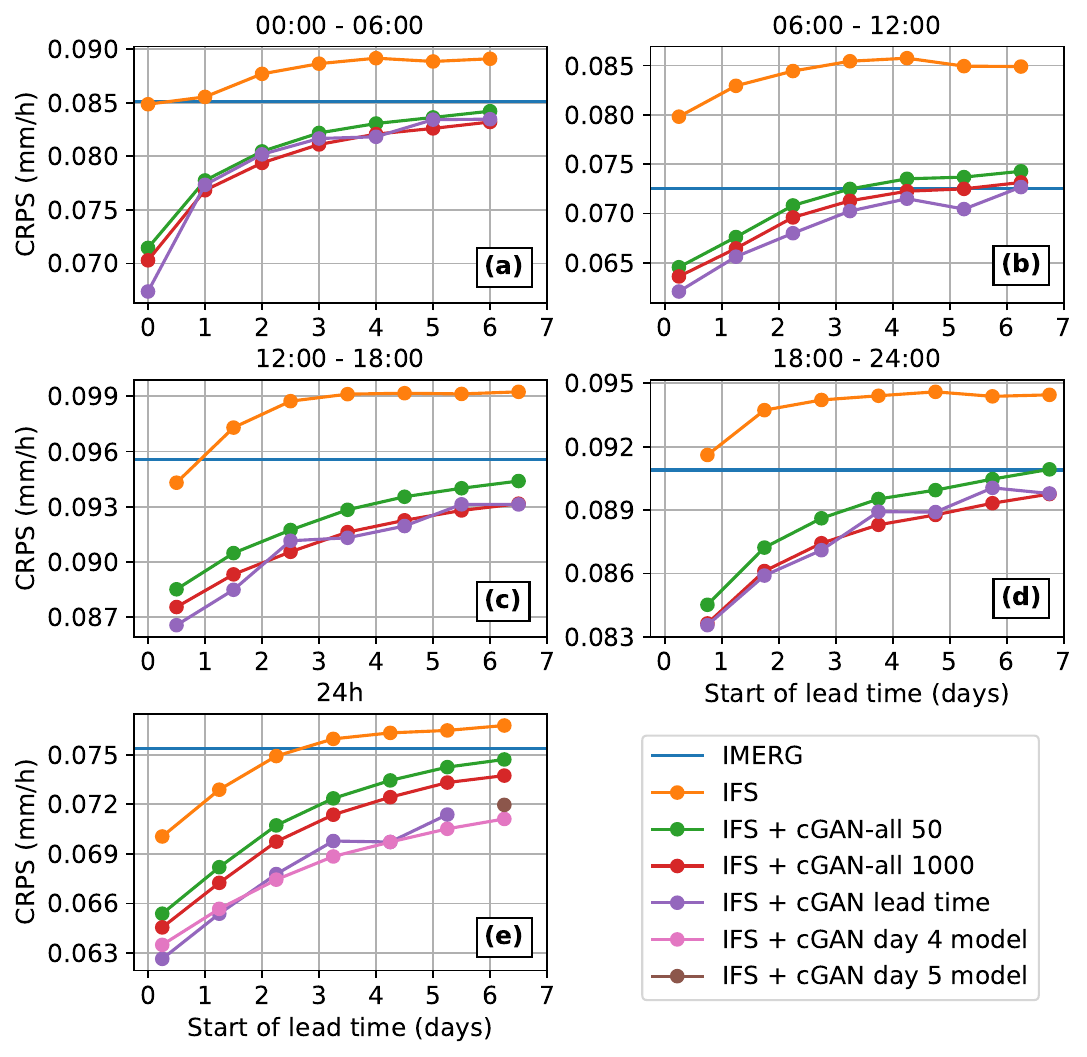}
\caption{6h and 24h rainfall accumulation, area and one-year time-mean CRPS. The top four plots (a,b,c,d) represent the four different 6-hour rainfall accumulation periods per day. Times are UTC. The bottom plot (e) represents rainfall accumulation from 06:00 to 06:00. All forecasts are initialised at 00:00 UTC. Each point represents the start of the accumulation period. The blue line is the CRPS of the IMERG climatological forecast. cGAN-all 50 denotes a 50 ensemble member forecast trained on all lead times. cGAN-all 1000 denotes the same but with 1000 ensemble members. cGAN lead time is the 1000-member forecast models trained separately on each lead time. In plot (e) the model trained on day 5 is used to forecast day 6 (brown point). The model trained on day 4 and applied to other days is also shown. Lower is better. For robustness, compare to Fig. S8.}
\label{fig:cGAN_progress}
\end{figure}

Initially 50-member ensemble forecasts were produced, consistent with the 50-member IFS ensemble, using a single cGAN model trained using all lead times. The CRPS was reduced by increasing the ensemble size to 1000 members and therefore producing a more accurate rainfall distribution, see Fig. \ref{fig:cGAN_progress}. Systematic errors in the IFS ensemble distribution change with forecast lead time, as different physical processes exert influences on different timescales. Rather than attempt to capture this in a single model the area mean CRPS was reduced further in some cases by training multiple models, each only predicting, and using data from, a single forecast lead time. For the 24h accumulation period, Fig. \ref{fig:cGAN_progress}e, the model trained on day 4 appeared to be a lucky outlier. This model applied to other days further reduced the CRPS. The final choice for 24h rainfall accumulation is then the models trained on lead times of 6h, 30h and 102h (4 days + 6 hours). For the 6h accumulation periods, Fig. \ref{fig:cGAN_progress}a,b,c,d, the training lead time always corresponds to the forecast lead time. These lead time dependent models were trained after we had seen the other cGAN results for the 2023-6-1 to 2024-6-1 test period, breaking a strict separation between model training and test data. However, all models were trained before we had seen the 2024-9-23 to 2025-9-23 test data, which is documented in the supporting information.

Quantile mapping was applied to cGAN outputs, for example in Fig. \ref{fig:QM}. Although aspects of the distribution were improved, skill measured by the area mean CRPS was worse, see Fig. S9.

\footnotesize
\section*{Author contributions}
Author contributions: F.C.C. performed the research and wrote the paper. A.T.T.M. adapted the GAN code from the model developed in \cite{Harris2022}. F.C.C., A.W., T.P. and J.M., guided the direction of the paper. A.T.T.M., S.N., B.A., Z.M., B.T. and B.K. provided data download and processing and valuable technical insight. F.P. and M.C. organised data access and compute resources. All authors provided valuable comments, suggestions and support for both production of the manuscript and direction of the research.

\footnotesize
\section*{Acknowledgements}
We acknowledge funding from Google.org via the United Nations World Food Programme. A. Weisheimer acknowledges funding from the National Centre for Atmospheric Science, UK.
The authors declare no competing interest.

\normalsize
\bibliography{pnas-sample}

@article {Ageet2022,
      author = "Simon Ageet and Andreas H. Fink and Marlon Maranan and Jeremy E. Diem and Joel Hartter and Andrew L. Ssali and Prosper Ayabagabo",
      title = "Validation of Satellite Rainfall Estimates over Equatorial {E}ast {A}frica",
      journal = "Journal of Hydrometeorology",
      year = "2022",
      publisher = "American Meteorological Society",
      address = "Boston MA, USA",
      volume = "23",
      number = "2",
      doi = "https://doi.org/10.1175/JHM-D-21-0145.1",
      pages=      "129 - 151",
      url = "https://journals.ametsoc.org/view/journals/hydr/23/2/JHM-D-21-0145.1.xml"
}

@article {Ageet2023,
      author = "Simon Ageet and Andreas H. Fink and Marlon Maranan and Benedikt Schulz",
      title = "Predictability of Rainfall over Equatorial East {A}frica in the {ECMWF} Ensemble Reforecasts on Short- to Medium-Range Time Scales",
      journal = "Weather and Forecasting",
      year = "2023",
      publisher = "American Meteorological Society",
      address = "Boston MA, USA",
      volume = "38",
      number = "12",
      doi = "https://doi.org/10.1175/WAF-D-23-0093.1",
      pages=      "2613 - 2630",
}

@misc{Alexe2024,
      title={GraphDOP: Towards skilful data-driven medium-range weather forecasts learnt and initialised directly from observations}, 
      author={Mihai Alexe and Eulalie Boucher and Peter Lean and Ewan Pinnington and Patrick Laloyaux and Anthony McNally and Simon Lang and Matthew Chantry and Chris Burrows and Marcin Chrust and Florian Pinault and Ethel Villeneuve and Niels Bormann and Sean Healy},
      year={2024},
      eprint={2412.15687},
      archivePrefix={arXiv},
      primaryClass={physics.ao-ph},
      note={https://doi.org/10.48550/arXiv.2412.15687}, 
}

@article{Alfieri2012,
  author = {Alfieri, Lorenzo and Thielen, Jutta and Pappenberger, Florian},
  title = {Operational early warning systems for water-related hazards in {E}urope},
  journal = {Environmental Science \& Policy},
  volume = {21},
  pages = {35--49},
  year = {2012},
  doi = {https://doi.org/10.1016/j.envsci.2012.01.008}
}

@article{Amell2025,
    author = {Amell, Adrià and Hee, Lilian and Pfreundschuh, Simon and Eriksson, Patrick},
    title = {Probabilistic Near-Real-Time Retrievals of Rain Over {A}frica Using Deep Learning},
    journal = {Journal of Geophysical Research: Atmospheres},
    volume = {130},
    number = {20},
    pages = {e2025JD044595},
    doi = {https://doi.org/10.1029/2025JD044595},
    year = {2025}
}

@article{Antonio2024,
    author = {Antonio, Bobby and McRae, Andrew T. T. and MacLeod, David and Cooper, Fenwick C. and Marsham, John and Aitchison, Laurence and Palmer, Tim N. and Watson, Peter A. G.},
    title = {Postprocessing East {A}frican Rainfall Forecasts Using a Generative Machine Learning Model},
    journal = {Journal of Advances in Modeling Earth Systems},
    volume = {17},
    number = {3},
    pages = {e2024MS004796},
    doi = {https://doi.org/10.1029/2024MS004796},
    year = {2025}
}

@InProceedings{Arjovsky2017,
  title = 	 {{W}asserstein Generative Adversarial Networks},
  author =       {Martin Arjovsky and Soumith Chintala and L{\'e}on Bottou},
  booktitle = 	 {Proceedings of the 34th International Conference on Machine Learning},
  pages = 	 {214--223},
  year = 	 {2017},
  editor = 	 {Precup, Doina and Teh, Yee Whye},
  volume = 	 {70},
  series = 	 {Proceedings of Machine Learning Research},
  month = 	 {06--11 Aug},
  publisher =    {PMLR},
  url = 	 {https://proceedings.mlr.press/v70/arjovsky17a.html},
}

@article {Arnold2024,
      author = "Sebastian Arnold and Eva-Maria Walz and Johanna Ziegel and Tilmann Gneiting",
      title = "Decompositions of the mean continuous ranked probability score.",
      journal = "Electronic Journal of Statistics",
      year = "2024",
      volume = "18",
      number = "2",
      doi = "https://doi.org/10.1214/24-EJS2316",
      pages= "4992--5044",
}

@article{Beck2018,
    author = {Beck, Hylke E. and Zimmermann Niklaus E. and McVicar, Tim R. and Vergopolan, Noemi and Berg, Alexis and Wood, Eric F.},
    title = {Present and future {K}öppen-{G}eiger climate classification maps at 1-km resolution},
    year = 2018,
    journal = {Scientific data},
    volume = 5,
    number = 180214,
    doi = {https://doi.org/10.1038/sdata.2018.214},
}

@article{Berthou2019,
    title = {Improved climatological precipitation characteristics over West Africa at convection-permitting scales.},
    author = {Berthou, Ségolène and Rowell, David P. and Kendon, Elizabeth J. and Roberts, Malcolm J. and Stratton, Rachel A. and Crook, Julia A. and Wilcox, Catherine},
    journal = {Climate Dynamics},
    volume = {53},
    pages = {1991-2011},
    year = {2019},
    doi = {https://doi.org/10.1007/s00382-019-04759-4}
}

@article{Bouallegue2023,
  title={Statistical modeling of 2-m temperature and 10-m wind speed forecast errors},
  author={Ben Bouallègue, Zied and Cooper, Fenwick and Chantry, Matthew and Düben, Peter and Bechtold, Peter and Sandu, Irina},
  journal={Monthly Weather Review},
  volume={151},
  number={4},
  pages={897--911},
  year={2023},
  doi={https://doi.org/10.1175/MWR-D-22-0107.1},
}

@article{Candille2005,
    author = {Candille, G. and Talagrand, O.},
    title = {Evaluation of probabilistic prediction systems for a scalar variable},
    journal = {Quarterly Journal of the Royal Meteorological Society},
    volume = {131},
    number = {609},
    pages = {2131-2150},
    doi = {https://doi.org/10.1256/qj.04.71},
    year = {2005}
}

@article{Chen2023,
  title={Fu{X}i: a cascade machine learning forecasting system for 15-day global weather forecast.},
  author={Chen, Lei and Zhong, Xiaohui and Zhang, Feng and Cheng, Yuan and Xu, Yinghui and Qi, Yuan and Li, Hao},
  journal={npj Climate and Atmospheric Science},
  volume={6},
  number={190},
  year={2023},
  doi={https://doi.org/10.1038/s41612-023-00512-1},
}

@article{Demeritt2013,
  author = {Demeritt, David and Nobert, Sébastien and Cloke, Hannah L. and Pappenberger, Florian},
  title = {The {E}uropean Flood Alert System and the communication, perception, and use of ensemble predictions for operational flood risk management},
  journal = {Hydrological Processes},
  volume = {27},
  number = {1},
  pages = {147--157},
  year = {2013},
  doi = {https://doi.org/10.1002/hyp.9419}
}

@article {Endris2021,
      author = "Hussen Seid Endris and Linda Hirons and Zewdu Tessema Segele and Masilin Gudoshava and Steve Woolnough and Guleid A. Artan",
      title = "Evaluation of the Skill of Monthly Precipitation Forecasts from Global Prediction Systems over the Greater Horn of {A}frica",
      journal = "Weather and Forecasting",
      year = "2021",
      publisher = "American Meteorological Society",
      address = "Boston MA, USA",
      volume = "36",
      number = "4",
      doi = "https://doi.org/10.1175/WAF-D-20-0177.1",
      pages=      "1275 - 1298",
}

@techreport{ESAWorldCover2020,
	title = "{ESA WorldCover} 10 m 2020 v100",
	author = "Zanaga, D. and Van De Kerchove, R. and De Keersmaecker, W. and Souverijns, N. and Brockmann, C. and Quast, R. and Wevers, J. and Grosu, A. and Paccini, A. and Vergnaud, S. and Cartus, O. and Santoro, M. and Fritz, S. and Georgieva, I. and Lesiv, M. and Carter, S. and Herold, M. and Li, Linlin and Tsendbazar, N. E. and Ramoino, F. and Arino, O.",
	institution = "European Space Agency (ESA)",
	year = 2021,
	doi = {https://doi.org/10.5281/zenodo.5571936},
	url = {https://esa-worldcover.org/}
}

@misc{FASTA,
  title = {{F}orecasting {A}frican {ST}orms {A}pplication ({FASTA})},
  howpublished = {\url{https://fastaweather.com}},
  year = 2026,
  note = {Accessed: 2026-10-8}
}

@article{Ferro2008,
    author = {Ferro, Christopher A. T. and Richardson, David S. and Weigel, Andreas P.},
    title = {On the effect of ensemble size on the discrete and continuous ranked probability scores},
    journal = {Meteorological Applications},
    volume = {15},
    number = {1},
    pages = {19-24},
    doi = {https://doi.org/10.1002/met.45},
    year = {2008}
}

@article {Funk2022,
      author = "Chris C. Funk and Pete Peterson and George J. Huffman and Martin Francis Landsfeld and Christa Peters-Lidard and Frank Davenport and Shraddhanand Shukla and Seth Peterson and Diego H. Pedreros and Alex C. Ruane and Carolyn Mutter and Will Turner and Laura Harrison and Austin Sonnier and Juliet Way-Henthorne and Gregory J. Husak",
      title = "Introducing and Evaluating the Climate Hazards Center {IMERG} with Stations (CHIMES): Timely Station-Enhanced Integrated Multisatellite Retrievals for Global Precipitation Measurement",
      journal = "Bulletin of the American Meteorological Society",
      year = "2022",
      publisher = "American Meteorological Society",
      address = "Boston MA, USA",
      volume = "103",
      number = "2",
      doi = "https://doi.org/10.1175/BAMS-D-20-0245.1",
      pages=      "E429 - E454",
}

@techreport{GMTED2010,
  title       = "Global multi-resolution terrain elevation data 2010 ({GMTED2010})",
  author      = "Danielson, J. J. and Gesch, D. B.",
  institution = "U.S. Geological Survey Open-File Report 2011–1073",
  year        = 2012,
}

@article{Gneiting2007,
    author = {Tilmann Gneiting and Adrian E Raftery},
    title = {Strictly Proper Scoring Rules, Prediction, and Estimation},
    journal = {Journal of the American Statistical Association},
    volume = {102},
    number = {477},
    pages = {359--378},
    year = {2007},
    doi = {https://doi.org/10.1198/016214506000001437},
}

@misc{Gneiting2025,
      title={Probabilistic measures afford fair comparisons of {AIWP} and {NWP} model output}, 
      author={Tilmann Gneiting and Tobias Biegert and Kristof Kraus and Eva-Maria Walz and Alexander I. Jordan and Sebastian Lerch},
      year={2025},
      eprint={2506.03744},
      archivePrefix={arXiv},
      primaryClass={stat.AP},
      note={https://doi.org/10.48550/arXiv.2506.03744}, 
}

@article{Gudoshava2022,
    author = {Gudoshava, Masilin and Wainwright, Caroline and Hirons, Linda and Endris, Hussen S. and Segele, Zewdu T. and Woolnough, Steve and Atheru, Zachary and Artan, Guleid},
    title = {Atmospheric and oceanic conditions associated with early and late onset for Eastern {A}frica short rains},
    journal = {International Journal of Climatology},
    volume = {42},
    number = {12},
    pages = {6562-6578},
    doi = {https://doi.org/10.1002/joc.7627},
    year = {2022}
}

@article{Gudoshava2024,
    author = {Gudoshava, Masilin and Nyinguro, Patricia and Talib, Joshua and Wainwright, Caroline and Mwanthi, Anthony and Hirons, Linda and de Andrade, Felipe and Mutemi, Joseph and Gitau, Wilson and Thompson, Elisabeth and Gacheru, Jemimah and Marsham, John and Endris, Hussen Seid and Woolnough, Steven and Segele, Zewdu and Atheru, Zachary and Artan, Guleid},
    title = {Drivers of sub-seasonal extreme rainfall and their representation in {ECMWF} forecasts during the Eastern {A}frican March-to-May seasons of 2018–2020},
    journal = {Meteorological Applications},
    volume = {31},
    number = {5},
    pages = {e70000},
    doi = {https://doi.org/10.1002/met.70000},
    year = {2024}
}

@misc{Gulrajani2017,
      title={Improved Training of {W}asserstein {GAN}s}, 
      author={Ishaan Gulrajani and Faruk Ahmed and Martin Arjovsky and Vincent Dumoulin and Aaron Courville},
      year={2017},
      eprint={1704.00028},
      archivePrefix={arXiv},
      primaryClass={cs.LG},
      note={https://doi.org/10.48550/arXiv.1704.00028}, 
}

@article{Harris2022,
	author = {Harris, Lucy and McRae, Andrew T. T. and Chantry, Matthew and Dueben, Peter D. and Palmer, Tim N.},
	title = {A Generative Deep Learning Approach to Stochastic Downscaling of Precipitation Forecasts},
	journal = {Journal of Advances in Modeling Earth Systems},
	volume = {14},
	number = {10},
	doi = {https://doi.org/10.1029/2022MS003120},
	year = {2022}
}

@mastersthesis{Hee2022,
  author  = "Lilian Hee",
  title   = "Rain over {A}frica. An application of quantile regression neural networks to retrieve precipitation from geostationary satellites",
  school  = "Chalmers university of technology, Gothenburg, Sweden",
  year    = "2022",
  note    = "https://hdl.handle.net/20.500.12380/305472",
}

@article{Henzi2021,
    author = {Henzi, Alexander and Ziegel, Johanna F. and Gneiting, Tilmann},
    title = {Isotonic Distributional Regression},
    journal = {Journal of the Royal Statistical Society Series B: Statistical Methodology},
    volume = {83},
    number = {5},
    pages = {963-993},
    year = {2021},
    month = {08},
    doi = {https://doi.org/10.1111/rssb.12450},
}

@article{Hersbach2020,
    author = {Hersbach, Hans and Bell, Bill and Berrisford, Paul and Hirahara, Shoji and Horányi, András and Muñoz-Sabater, Joaquín and Nicolas, Julien and Peubey, Carole and Radu, Raluca and Schepers, Dinand and Simmons, Adrian and Soci, Cornel and Abdalla, Saleh and Abellan, Xavier and Balsamo, Gianpaolo and Bechtold, Peter and Biavati, Gionata and Bidlot, Jean and Bonavita, Massimo and De Chiara, Giovanna and Dahlgren, Per and Dee, Dick and Diamantakis, Michail and Dragani, Rossana and Flemming, Johannes and Forbes, Richard and Fuentes, Manuel and Geer, Alan and Haimberger, Leo and Healy, Sean and Hogan, Robin J. and Hólm, Elías and Janisková, Marta and Keeley, Sarah and Laloyaux, Patrick and Lopez, Philippe and Lupu, Cristina and Radnoti, Gabor and de Rosnay, Patricia and Rozum, Iryna and Vamborg, Freja and Villaume, Sebastien and Thépaut, Jean-Noël},
    title = {The {ERA}5 global reanalysis},
    journal = {Quarterly Journal of the Royal Meteorological Society},
    volume = {146},
    number = {730},
    pages = {1999-2049},
    doi = {https://doi.org/10.1002/qj.3803},
    year = {2020}
}

@Inbook{Huffman2020,
    author="Huffman, George J. and Bolvin, David T. and Braithwaite, Dan and Hsu, Kuo-Lin and Joyce, Robert J. and Kidd, Christopher and Nelkin, Eric J. and Sorooshian, Soroosh and Stocker, Erich F. and Tan, Jackson and Wolff, David B. and Xie, Pingping",
    editor="Levizzani, Vincenzo and Kidd, Christopher and Kirschbaum, Dalia B. and Kummerow, Christian D. and Nakamura, Kenji and Turk, F. Joseph",
    title="Integrated Multi-satellite Retrievals for the Global Precipitation Measurement ({GPM}) Mission ({IMERG})",
    bookTitle="Satellite Precipitation Measurement: Volume 1",
    year="2020",
    publisher="Springer International Publishing",
    address="Cham",
    pages="343--353",
    doi="https://doi.org/10.1007/978-3-030-24568-9_19",
}

@techreport{Huffman2023,
    author      = "Huffman, George J. and Bolvin, David T. and Joyce, Robert and Nelkin, Eric J. and Tan, Jackson and Braithwaite, Dan and Hsu, Kuolin and Kelley, Owen A. and Nguyen, Phu and Sorooshian, Soroosh and Watters, Daniel C. and West, B. Jason and Xie, Pingping",
    title       = "{NASA} Global Precipitation Measurement ({GPM}) Integrated Multi-satellitE Retrievals for {GPM} ({IMERG}) Version 07",
    institution = "National Aeronautics and Space Administration (NASA)",
    address     = "Code 612 Greenbelt, MD 20771",
    year        = 2023,
    url         = {https://gpm.nasa.gov/data/imerg}
}

@book{IFS,
	author    = "ECMWF",
    title     = "IFS Documentation CY49R1 - Part V: Ensemble Prediction System",
	year      = "2024",
	note = {https://doi.org/10.21957/956d60ad81}
}

@article{Kolstad2024,
    author = {Kolstad, Erik W. and Parker, Douglas J. and MacLeod, David A. and Wainwright, Caroline M. and Hirons, Linda C.},
    title = {Beyond the regional average: Drivers of geographical rainfall variability during East {A}frica's short rains},
    journal = {Quarterly Journal of the Royal Meteorological Society},
    volume = {150},
    number = {764},
    pages = {4550-4566},
    doi = {https://doi.org/10.1002/qj.4829},
    year = {2024}
}

@article{Lam2023,
    author = {Remi Lam  and Alvaro Sanchez-Gonzalez  and Matthew Willson  and Peter Wirnsberger  and Meire Fortunato  and Ferran Alet  and Suman Ravuri  and Timo Ewalds  and Zach Eaton-Rosen  and Weihua Hu  and Alexander Merose  and Stephan Hoyer  and George Holland  and Oriol Vinyals  and Jacklynn Stott  and Alexander Pritzel  and Shakir Mohamed  and Peter Battaglia },
    title = {Learning skillful medium-range global weather forecasting},
    journal = {Science},
    volume = {382},
    number = {6677},
    pages = {1416-1421},
    year = {2023},
    doi = {https://doi.org/10.1126/science.adi2336},
}

@misc{Lang2024a,
      title={{AIFS} -- {ECMWF}'s data-driven forecasting system}, 
      author={Simon Lang and Mihai Alexe and Matthew Chantry and Jesper Dramsch and Florian Pinault and Baudouin Raoult and Mariana C. A. Clare and Christian Lessig and Michael Maier-Gerber and Linus Magnusson and Zied Ben Bouallègue and Ana Prieto Nemesio and Peter D. Dueben and Andrew Brown and Florian Pappenberger and Florence Rabier},
      year={2024},
      eprint={2406.01465},
      archivePrefix={arXiv},
      primaryClass={physics.ao-ph},
      note={https://doi.org/10.48550/arXiv.2406.01465}, 
}

@misc{Lang2024b,
      title={{AIFS-CRPS}: Ensemble forecasting using a model trained with a loss function based on the Continuous Ranked Probability Score}, 
      author={Simon Lang and Mihai Alexe and Mariana C. A. Clare and Christopher Roberts and Rilwan Adewoyin and Zied Ben Bouallègue and Matthew Chantry and Jesper Dramsch and Peter D. Dueben and Sara Hahner and Pedro Maciel and Ana Prieto-Nemesio and Cathal O'Brien and Florian Pinault and Jan Polster and Baudouin Raoult and Steffen Tietsche and Martin Leutbecher},
      year={2024},
      eprint={2412.15832},
      archivePrefix={arXiv},
      primaryClass={physics.ao-ph},
      note={https://doi.org/10.48550/arXiv.2412.15832}, 
}

@article {Lavers2021,
      author = "David A. Lavers and Shaun Harrigan and Christel Prudhomme",
      title = "Precipitation Biases in the {ECMWF} Integrated Forecasting System",
      journal = "Journal of Hydrometeorology",
      year = "2021",
      publisher = "American Meteorological Society",
      address = "Boston MA, USA",
      volume = "22",
      number = "5",
      doi = "https://doi.org/10.1175/JHM-D-20-0308.1",
      pages=      "1187 - 1198"
}

@misc{Leinonen2023,
      title={Latent diffusion models for generative precipitation nowcasting with accurate uncertainty quantification}, 
      author={Jussi Leinonen and Ulrich Hamann and Daniele Nerini and Urs Germann and Gabriele Franch},
      year={2023},
      eprint={2304.12891},
      archivePrefix={arXiv},
      primaryClass={physics.ao-ph},
      note = {https://doi.org/10.48550/arXiv.2304.12891}, 
}

@article{Lerch2017,
    address = {Hayward, Calif :},
    issn = {0883-4237},
    journal = {Statistical science : a review journal of the Institute of Mathematical Statistics},
    lccn = {LC 86-642668},
    number = {1},
    publisher = {Institute of Mathematical Statistics},
    title = {Forecaster’s Dilemma: Extreme Events and Forecast Evaluation},
    volume = {32},
    year = {2017-2-1},
    author = {Lerch, Sebastian and Thorarinsdottir, Thordis L and Ravazzolo, Francesco and Gneiting, Tilmann},
    url = "https://www.jstor.org/stable/26408123",
}

@article{Leutbecher2019,
    author = {Leutbecher, Martin},
    title = {Ensemble size: How suboptimal is less than infinity?},
    journal = {Quarterly Journal of the Royal Meteorological Society},
    volume = {145},
    number = {S1},
    pages = {107-128},
    doi = {https://doi.org/10.1002/qj.3387},
    year = {2019}
}

@article{Li2024,
    author = {Lizao Li  and Robert Carver  and Ignacio Lopez-Gomez  and Fei Sha  and John Anderson},
    title = {Generative emulation of weather forecast ensembles with diffusion models},
    journal = {Science Advances},
    volume = {10},
    number = {13},
    pages = {eadk4489},
    year = {2024},
    doi = {https://doi.org/10.1126/sciadv.adk4489},
}

@inbook{Maraun2018,
    place={Cambridge},
    title={Statistical Downscaling Concepts and Methods},
    booktitle={Statistical Downscaling and Bias Correction for Climate Research},
    publisher={Cambridge University Press},
    author={Maraun, Douglas and Widmann, Martin},
    year={2018},
    pages={133–134}
}

@article{Mardani2025,
    author = {Mardani, Morteza and Brenowitz, Noah and Cohen, Yair and Pathak, Jaideep and Chen, Chieh-Yu and Liu, Cheng-Chin and Vahdat, Arash and Nabian, Mohammad Amin and Ge, Tao and Subramaniam, Akshay and Kashinath, Karthik and Kautz, Jan and Pritchard, Mike},
    title = {Residual corrective diffusion modeling for km-scale atmospheric downscaling},
    journal = {Communications Earth \& Environment},
    volume = {6},
    pages = {1214},
    year = {2025},
    doi = {https://doi.org/10.1038/s43247-025-02042-5}
}

@article{Mwanthi2024,
    author = {Mwanthi, Anthony M.  and Mutemi, Joseph N.  and Opijah, Franklin J.  and Mutua, Francis M.  and Atheru, Zachary  and Artan, Guleid },
    title = {Implications of WRF model resolutions on resolving rainfall variability with topography over East {A}frica},
    journal = {Frontiers in Climate},
    volume = {6},
    year = {2024},
    doi = {https://doi.org/10.3389/fclim.2024.1311088}
}

@article{Nath2026,
    author = {Shruti Nath and David Koros and Fenwick C. Cooper and David MacLeod and Hannah Kimani and Zacharia Mwai and Asaminew Teshome and Masilin Gudoshava and Isaac Obai and Maurine Ambani and Mark Arango and Jesse Mason and Matthew Chantry and Florian Pappenberger and Antje Weisheimer and Tim Palmer},
    title = {Predicting {E}ast {A}frica's rainfall extremes with calibrated {AI} post-processing systems},
    journal = {{NPJ} Natural Hazards},
    year = {2026},
    note = {Accepted for publication. Preprint: https://doi.org/10.21203/rs.3.rs-10487940/v1}
}

@article{Nicholson2017,
    author = {Nicholson, Sharon E.},
    title = {Climate and climatic variability of rainfall over eastern {A}frica},
    journal = {Reviews of Geophysics},
    volume = {55},
    number = {3},
    pages = {590-635},
    doi = {https://doi.org/10.1002/2016RG000544},
    year = {2017}
}

@article{Omay2025,
	author = {Omay, Paulino Omoj and Muthama, Nzioka J. and Oludhe, Christopher and Kinama, Josiah M. and Artan, Guleid and Atheru, Zachary},
	title = {Evaluation of satellite-based rainfall estimates over the {IGAD} region of Eastern {A}frica},
	journal = {Meteorology and Atmospheric Physics},
    volume = {137},
    pages = {22},
	year = 2025,
    doi = {https://doi.org/10.1007/s00703-025-01068-w}
}

@article{Palmer2023,
	author = {Palmer, Paul I. and Wainwright, Caroline M. and Dong, Bo and Maidment, Ross I. and Wheeler, Kevin G. and Gedney, Nicola and Hickman, Jonathan E. and Madani, Nima and Folwell, Sonja S. and Abdo, Gamal and Allan, Richard P. and Black, Emily C. L. and Feng, Liang and Gudoshava, Masilin and Haines, Keith and Huntingford, Chris and Kilavi, Mary and Lunt, Mark F. and Shaaban, Ahmed and Turner, Andrew G.},
	title = {Drivers and impacts of Eastern {A}frican rainfall variability},
	journal = {Nature Reviews Earth \& Environment},
    volume = {4},
    pages = {254--270},
	year = 2023,
    doi = {https://doi.org/10.1038/s43017-023-00397-x}
}

@Article{Papacharalampous2023,
    AUTHOR = {Papacharalampous, Georgia and Tyralis, Hristos and Doulamis, Anastasios and Doulamis, Nikolaos},
    TITLE = {Comparison of Machine Learning Algorithms for Merging Gridded Satellite and Earth-Observed Precipitation Data},
    JOURNAL = {Water},
    VOLUME = {15},
    YEAR = {2023},
    NUMBER = {4},
    ARTICLE-NUMBER = {634},
    DOI = {https://doi.org/10.3390/w15040634}
}

@article{Price2024,
	author = {Price, Ilan and Sanchez-Gonzalez, Alvaro and Alet, Ferran and Andersson, Tom R. and El-Kadi, Andrew and Masters, Dominic and Ewalds, Timo and Stott, Jacklynn and Mohamed, Shakir and Battaglia, Peter and Lam, Remi and Willson, Matthew},
	title = {Probabilistic weather forecasting with machine learning},
	journal = {Nature},
    volume = {637},
    pages = {84--90},
	year = 2024,
    doi = {https://doi.org/10.1038/s41586-024-08252-9}
}

@article{Rajani2022,
    title = {Review of {GPM} {IMERG} performance: A global perspective},
    journal = {Remote Sensing of Environment},
    volume = {268},
    pages = {112754},
    year = {2022},
    doi = {https://doi.org/10.1016/j.rse.2021.112754},
    author = {Rajani K. Pradhan and Yannis Markonis and Mijael Rodrigo {Vargas Godoy} and Anahí Villalba-Pradas and Konstantinos M. Andreadis and Efthymios I. Nikolopoulos and Simon Michael Papalexiou and Akif Rahim and Francisco J. Tapiador and Martin Hanel}
}

@article{Roberts2022,
    author = {Roberts, Alexander J. and Fletcher, Jennifer K. and Groves, James and Marsham, John H. and Parker, Douglas J. and Blyth, Alan M. and Adefisan, Elijah A. and Ajayi, Vincent O. and Barrette, Ronald and de Coning, Estelle and Dione, Cheikh and Diop, Abdoulahat and Foamouhoue, Andre K. and Gijben, Morne and Hill, Peter G. and Lawal, Kamoru A. and Mutemi, Joseph and Padi, Michael and Popoola, Temidayo I. and Rípodas, Pilar and Stein, Thorwald H.M. and Woodhams, Beth J.},
    title = {Nowcasting for {A}frica: {A}dvances, potential and value},
    journal = {Weather},
    volume = {77},
    number = {7},
    pages = {250-256},
    doi = {https://doi.org/10.1002/wea.3936},
    year = {2022}
}

@article {Saha2014,
      author = "Suranjana Saha and Shrinivas Moorthi and Xingren Wu and Jiande Wang and Sudhir Nadiga and Patrick Tripp and David Behringer and Yu-Tai Hou and Hui-ya Chuang and Mark Iredell and Michael Ek and Jesse Meng and Rongqian Yang and Malaquías Peña Mendez and Huug van den Dool and Qin Zhang and Wanqiu Wang and Mingyue Chen and Emily Becker",
      title = "The {NCEP} Climate Forecast System Version 2",
      journal = "Journal of Climate",
      year = "2014",
      publisher = "American Meteorological Society",
      address = "Boston MA, USA",
      volume = "27",
      number = "6",
      doi = "https://doi.org/10.1175/JCLI-D-12-00823.1",
      pages=      "2185 - 2208",
}

@techreport{Skamarock2005,
	author = {Skamarock, W. C. and Klemp, J. B. and Dudhia, J. and Gill, D. O. and Barker, D. M. and Wang, W. and Powers, J. G.},
	title = {A description of the Advanced Research {WRF} Version 2.},
	institution = {NCAR},
    type = {Tech. Note},
    number = {NCAR/TN-468+STR},
	year = 2005,
    doi = {https://doi.org/doi:10.5065/D6DZ069T}
}

@techreport{Skamarock2008,
	author = {Skamarock, W. C. and Klemp, J. B. and Dudhia, J. and Gill, D. O. and Barker, D. M. and Wang, W. and Powers, J. G.},
	title = {A Description of the Advanced Research {WRF} Version 3.},
	institution = {NCAR},
    type = {Tech. Note},
    number = {NCAR/TN-475+STR},
	year = 2008,
    doi = {https://doi.org/doi:10.5065/D68S4MVH}
}

@article{Stellingwerf2021,
    author = {Stellingwerf, Sippora and Riddle, Emily and Hopson, Thomas M. and Knievel, Jason C. and Brown, Barbara and Gebremichael, Mekonnen},
    title = {Optimizing Precipitation Forecasts for Hydrological Catchments in {E}thiopia Using Statistical Bias Correction and Multi-Modeling},
    journal = {Earth and Space Science},
    volume = {8},
    number = {6},
    pages = {e2019EA000933},
    doi = {https://doi.org/10.1029/2019EA000933},
    year = {2021}
}

@article {Vannitsem2021,
      author = "Stéphane Vannitsem and John Bjørnar Bremnes and Jonathan Demaeyer and Gavin R. Evans and Jonathan Flowerdew and Stephan Hemri and Sebastian Lerch and Nigel Roberts and Susanne Theis and Aitor Atencia and Zied Ben Bouallègue and Jonas Bhend and Markus Dabernig and Lesley De Cruz and Leila Hieta and Olivier Mestre and Lionel Moret and Iris Odak Plenković and Maurice Schmeits and Maxime Taillardat and Joris Van den Bergh and Bert Van Schaeybroeck and Kirien Whan and Jussi Ylhaisi",
      title = "Statistical Postprocessing for Weather Forecasts: Review, Challenges, and Avenues in a Big Data World",
      journal = "Bulletin of the American Meteorological Society",
      year = "2021",
      publisher = "American Meteorological Society",
      address = "Boston MA, USA",
      volume = "102",
      number = "3",
      doi = "https://doi.org/10.1175/BAMS-D-19-0308.1",
      pages=      "E681 - E699",
}

@article {Vogel2020,
      author = "Peter Vogel and Peter Knippertz and Andreas H. Fink and Andreas Schlueter and Tilmann Gneiting",
      title = "Skill of Global Raw and Postprocessed Ensemble Predictions of Rainfall in the Tropics",
      journal = "Weather and Forecasting",
      year = "2020",
      publisher = "American Meteorological Society",
      address = "Boston MA, USA",
      volume = "35",
      number = "6",
      doi = "https://doi.org/10.1175/WAF-D-20-0082.1",
      pages=      "2367 - 2385",
}

@article{Walz2024a,
	author = {Walz, Eva-Maria and Henzi, Alexander and Ziegel, Johanna and Gneiting, Tilmann},
	title = {Easy Uncertainty Quantification ({EasyUQ}): Generating Predictive Distributions from Single-Valued Model Output},
	journal = {SIAM Review},
	volume = {66},
	number = {1},
	pages = {91-122},
	year = {2024},
	doi = {https://doi.org/10.1137/22M1541915}
}

@article{Walz2024b,
	author = {Walz, Eva-Maria and Knippertz, Peter and Fink, Andreas H. and Köhler, Gregor and Gneiting, Tilmann},
	title = {Physics-Based vs Data-Driven 24-Hour Probabilistic Forecasts of Precipitation for Northern Tropical {A}frica},
	journal = {Monthly Weather Review},
	volume = {152},
	number = {9},
	pages = {2011–2031},
	year = {2024},
	doi = {https://doi.org/10.1175/MWR-D-24-0005.1}
}

@misc{WMO2018,
      title={Manual on Codes ({WMO}-{N}o. 306)}, 
      volume = {2},
      author = "{W}orld {M}eteorological {O}rganization",
      address = "Geneva",
      year={2018},
      note={https://library.wmo.int/idurl/4/35717}, 
}

@article{Xiong2025,
    title = {Continental evaluation of {GPM} {IMERG} {V07B} precipitation on a sub-daily scale},
    journal = {Remote Sensing of Environment},
    volume = {321},
    pages = {114690},
    year = {2025},
    doi = {https://doi.org/10.1016/j.rse.2025.114690},
    author = {Jinghua Xiong and Guoqiang Tang and Yuting Yang}
}

@misc{Zhang2025,
      title={Numerical models outperform {AI} weather forecasts of record-breaking extremes}, 
      author={Zhongwei Zhang and Erich Fischer and Jakob Zscheischler and Sebastian Engelke},
      year={2025},
      eprint={2508.15724},
      archivePrefix={arXiv},
      primaryClass={physics.ao-ph},
      note={https://doi.org/10.48550/arXiv.2508.15724}, 
}

@article{Zhong2025,
    author = {Xiaohui Zhong  and Lei Chen  and Hao Li  and Roberto Buizza  and Jun Liu  and Jie Feng  and Zijian Zhu  and Xu Fan  and Kan Dai  and Jing-jia Luo  and Jie Wu  and Bo Lu },
    title = {Fu{X}i-{ENS}: A machine learning model for efficient and accurate ensemble weather prediction},
    journal = {Science Advances},
    volume = {11},
    number = {44},
    pages = {eadu2854},
    year = {2025},
    doi = {https://doi.org/10.1126/sciadv.adu2854},
}

\clearpage

\section*{Supporting Information}

A list of the models compared in this paper and which figures they appear in is given in table S1.

In the main paper the test period 1st June 2023 to 1st June 2024 is considered. The analysis is expanded here to add the additional test year 23rd September 2024 to 23rd September 2025 which was not available when the models were trained. Comparing the results allows us to gain some idea of how robust they are. Plots from the later test period, equivalent to the plots from the earlier period in the main paper, include Figs. S3, S6, S7 and S8.

The GPM Core Observatory, the key satellite in the constellation used for IMERG, underwent an orbit boost on 7–8 November 2023. The 23 September 2024 to 23 September 2025 test period lies entirely after the boost, whereas the 1 June 2023 to 1 June 2024 period straddles it. If the boost were materially affecting our conclusions, we would expect the relative performance of the models to differ between the two periods. It does not: as shown in Figs. S3, S6, S7 and S8, the ordering and approximate parity of the post-processed models is largely preserved across both periods, despite the absolute CRPS values differing because of the differing amounts of rainfall.

The IMERG 7 final data product ended in October 2025. Additional analysis to place these results in context is detailed in the following sections.

\subsection*{Seasons}

The pattern of climatological rainfall, which closely matches the CRPS is given in Fig. S1. The rainfall is broken down into seasons in Fig. S2 where the two test periods may also be compared. The increased rainfall in the earlier test period is therefore a plausible explanation for the higher CRPS in this period. The CRPS maps for the seasonal equivalent of Figs. 2 and S3 are presented in S10, S11, S12, S13. Similarities between the post-processed models are once again clear, and the CRPS differences from the climatological distribution appear to be somewhat aligned to where it is raining.

\subsection*{Countries}

As shown on the maps (e.g. Fig. 2), the quality of forecast depends upon the location, or country. This is not illustrated as a function of lead time (e.g. Fig. 4). To address this the CRPS over select countries in the region is averaged and plotted in Figs. S14 and S15. While the CRPS of the IMERG climatology varies substantially, the relative order of the models is quite robust.

\subsection*{ROC and AUC scores}

Many other metrics for evaluating the forecast are routinely applied by meteorologists. For example, to test that the forecast probabilities line up with what actually happened, the reliability diagram is often used. We use a histogram in Fig. 5 to summarise the marginal distribution, without going into too much complex detail. Metrics that quantify the quality of spatial patterns include the Fractions Skill Score (FSS), the Radially Averaged Log Spectral Distance (RALSD) and the Variogram score. While useful as an aid to intuition, we argue that the spatial structure is not the main focus of the paper, despite this being the main reason for the development of the cGAN model in the first place. We felt that the CRPS, being a generalisation of the Mean-Absolute-Error, is preferable in this probabilistic context to metrics such as the Mean-Squared-Error and bias. Focussing variants of the CRPS on the tails of the distribution can be useful and is an active area of research, however the additional complexity is beyond our scope.

In the context of operational forecasting, relating a probabilistic forecast to a rainfall event of a certain magnitude (threshold) is important. One metric is the Brier score, which is closely related to the CRPS. A more established metric widely used by NHMS is the Receiver Operating Characteristic (ROC) curve. The ROC curve assesses the skill of a forecast exceeding a given threshold, by plotting the true positive rate (sensitivity) against the false positive rate (1—specificity), across the range of probability thresholds. See Fig. S16 for an example. The area under the ROC curve is a summary measure of how good the forecast is, giving rise to the Area Under the Curve (AUC) score. However, a higher AUC score does not necessarily mean that the model performs better at every decision threshold. The AUC score by country, season, test period and for a selection of important thresholds is given in Figs. S17 and S18.

\subsection*{Additional points}

Comprehensive evaluation of all of the possible forecast results, for both 6h and 24h accumulations of interest, is beyond the scope of this paper. Given the number of possible interesting features, more comprehensive evaluation is probably best tailored to the requirements a particular application. Nevertheless, there are some interesting additional points that we make here.

\begin{itemize}
\item We make a comparison to models without post-processing (Fig. S4) illustrating the substantial magnitude of the skill improvements.
\item To illustrate the maps of skill of the 6h accumulation forecasts and how they vary over the course of the day we include Fig. S5. The CRPS is improved during the 1200-1800 UTC convective period over Kenya, Somalia, Uganda and Tanzania. Rainfall over Lake Victoria and along the Indian Ocean coastline occurs at different times of the day and the corresponding CRPS is reduced. The failure of IFS+cGAN to beat climatology over the Congo and South Sudan in the 24h accumulation forecasts, Fig. 2 and S3, is also present in the 6h forecasts. The lack of robustness between years in the Ethiopian region is within the 1800-000 UTC and 0000-0600 UTC periods.
\item As shown in Fig. 5, application of quantile mapping corrects the rainfall distribution very well. However, naive application increases the CRPS of cGAN, making the forecast worse under this metric. See Fig. S9.
\end{itemize}

\begin{figure*}
\centering
\includegraphics[width=0.7\textwidth]{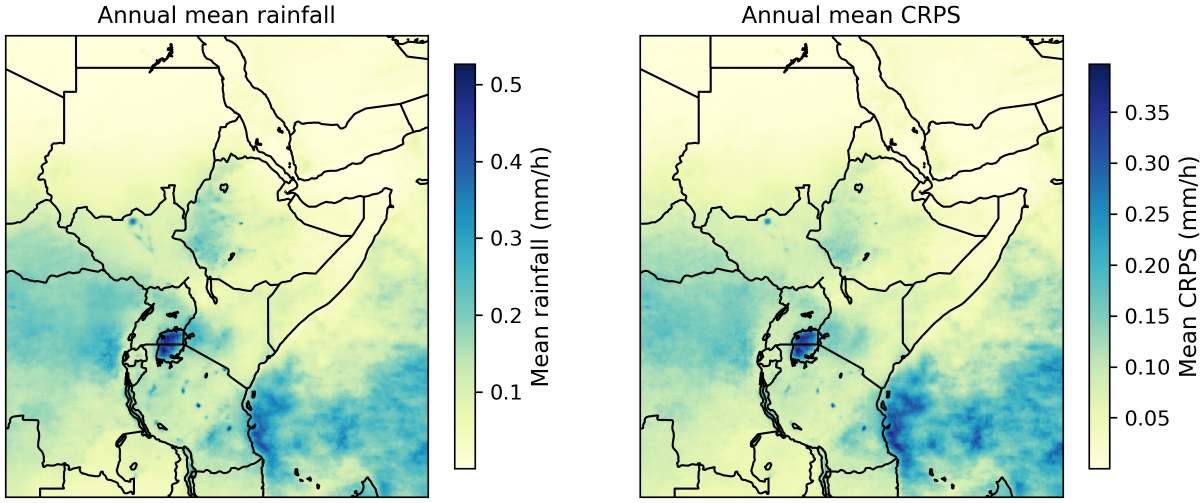}
\caption*{Figure S1: {\bf Left:} IMERG annual mean rainfall. {\bf Right:} Annual mean IMERG CRPS, both over the period 2023-6-1 to 2024-6-1. The rainfall is largest over and around Lake Victoria. The Congo rainforest, southern Indian Ocean and into Tanzania and the Ethiopian highlands are also areas of relatively high rainfall. The CRPS largely reflects where rainfall occurs and taking the area mean CRPS heavily weights these regions. Improvements to the CRPS in dry areas can only make a small contribution.}
\end{figure*}

\begin{figure*}
\centering
\includegraphics[width=\textwidth]{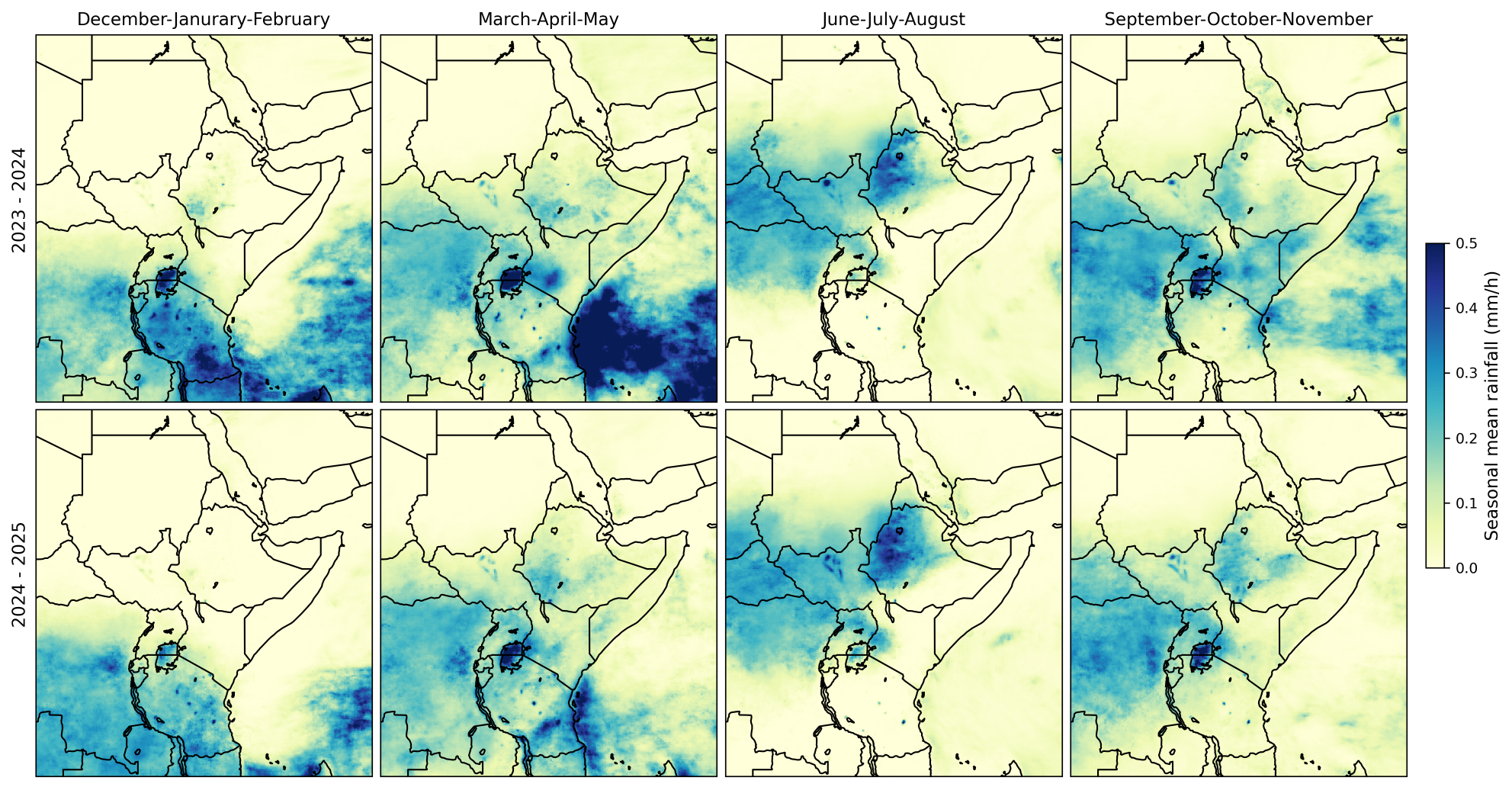}
\caption*{Figure S2: Mean rainfall during each season for the two test periods 2023-6-1 to 2024-6-1 (top) and 2024-9-23 to 2025-9-23 (bottom). There is less rain in the later test period. However, the spatial pattern of rainfall has similarities in both test periods, which is reflected in the similarity of the spatial patterns and relative improvements of forecast scores in both test periods. See Figs. S10, S11, S12 and S13.}
\end{figure*}

\begin{figure*}
\centering
\includegraphics[width=\textwidth]{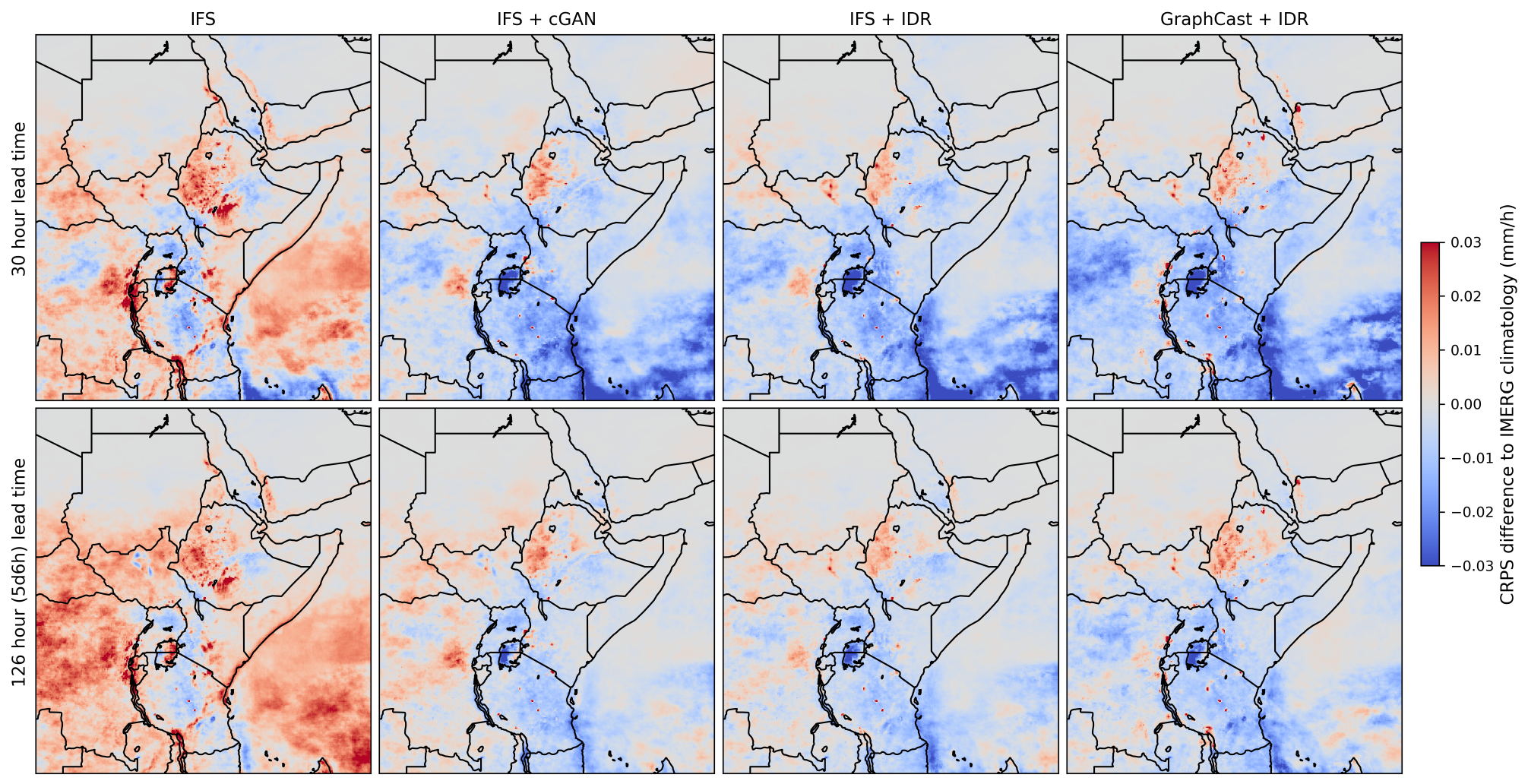}
\caption*{Figure S3: Difference between the CRPS of the 24h rainfall accumulation forecasts and the CRPS of the IMERG climatological distribution at two example lead times, 30h (top) and 126h (bottom), averaged over one year (2024-9-23 to 2025-9-23) for {\bf left}: IFS, {\bf middle left}: IFS with cGAN post-processing to 1000 ensemble members, {\bf middle} right: IFS with IDR post-processing and {\bf right}: GraphCast with IDR post-processing. Blue means that the model has a lower (better) CRPS. Red means that the IMERG climatological forecast has a lower CRPS. Compare to Fig. 2 in the main paper that is presented over a different period and with a different colour scale.}
\end{figure*}

\begin{figure*}
\centering
\includegraphics[width=0.875\textwidth]{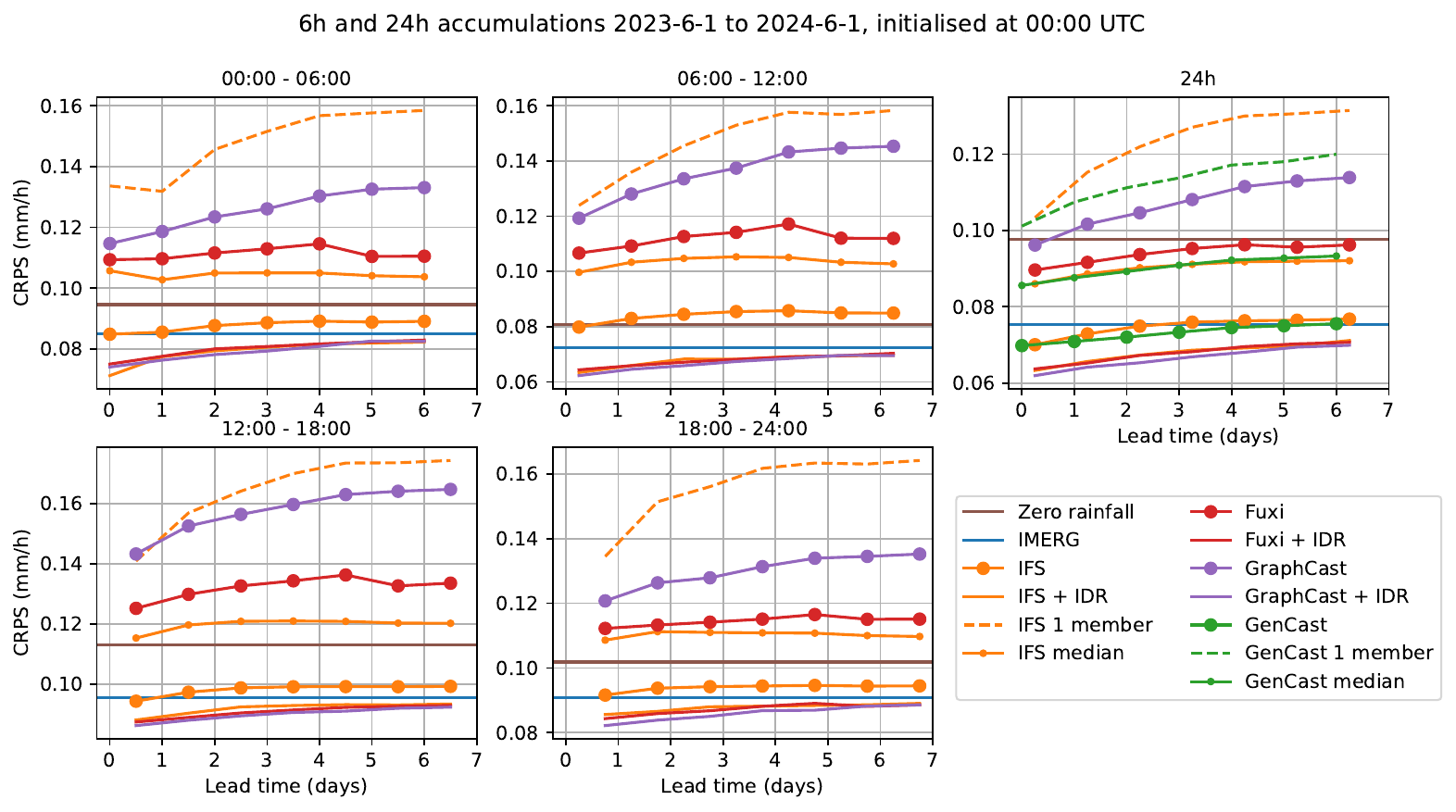}
\includegraphics[width=0.875\textwidth]{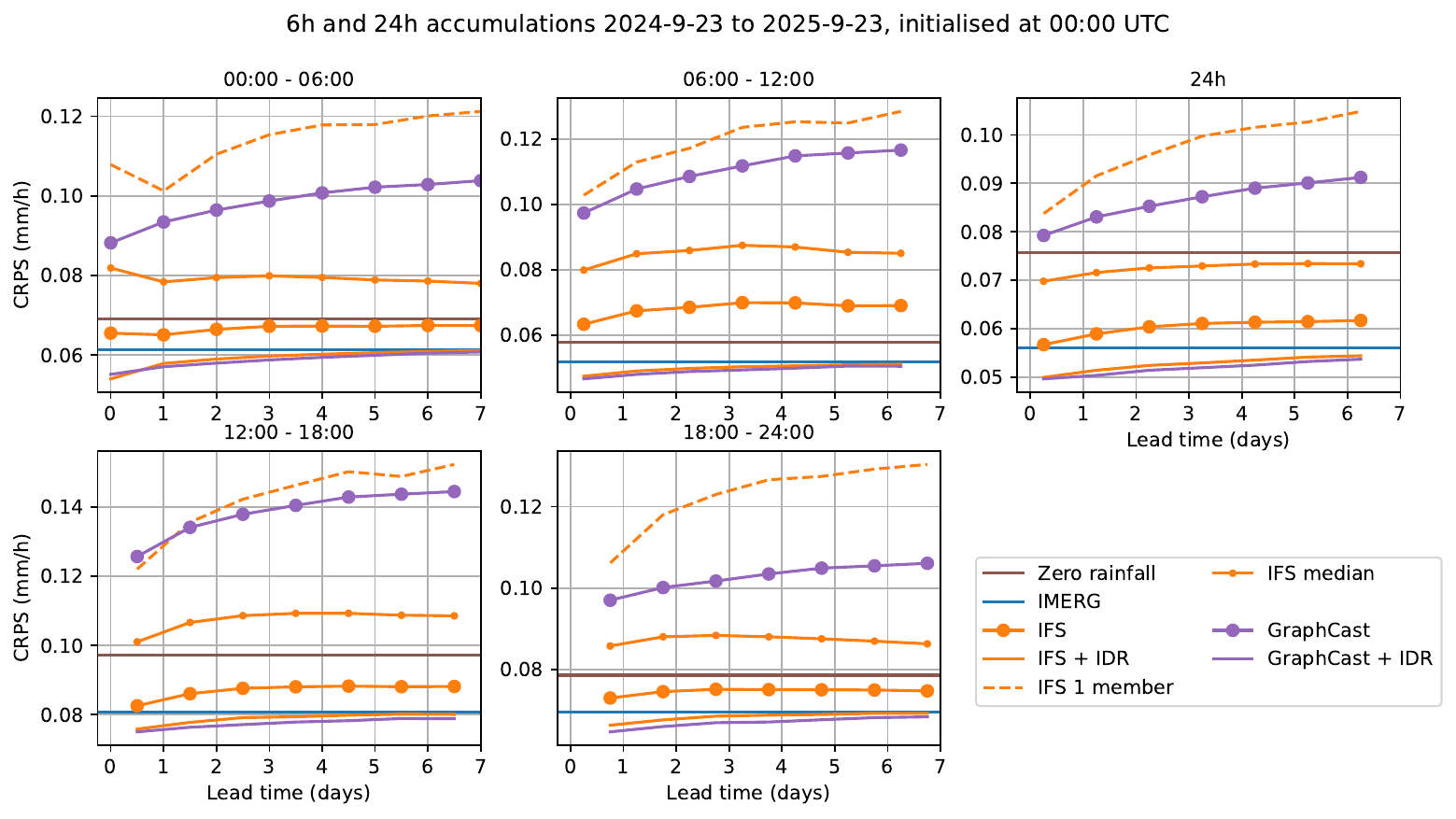}
\caption*{Figure S4: The CRPS of IFS, FuXi, GraphCast and GenCast without IDR applied for 6h rainfall accumulations at different times of the day and for 24h rainfall accumulations for two different test years, top 5 (2023-6-1 to 2024-6-1) and bottom 5 (2024-9-23 to 2025-9-23) plots). GenCast forecasts are only available with 12 hour accumulation periods starting at 00:00 and 12:00. The CRPS of predicting zero rainfall is given by the brown line. The blue line is the CRPS of the IMERG climatological distribution. For a deterministic forecast, a single ensemble member, the CRPS reduces to the mean absolute error. Lower is better. \\
\indent \hspace{3mm}
The deterministic FuXi and GraphCast models do better than the first ensemble member from the IFS forecast. However they are beaten by the IFS ensemble median which can be taken to be a deterministic forecast benchmark. The large CRPS might reflect to some extent that FuXi and GraphCast are trained on ERA5. However, so is GenCast and an IFS analysis variant is used to produce ERA5. GenCast performs marginally better than IFS. The GenCast ensemble contains 56 members in 2023 and 52 members in 2024. Sufficient historical data was not available to train post-processing of GenCast with either cGAN or IDR. \\
\indent \hspace{3mm}
Comparing the top 5 plots, earlier test period, and the bottom 5 plots, later test period, indicates different average CRPS over each period due to different numbers, locations and intensities of rainfall events. The earlier period had more rainfall (see Fig. S2), reflected in the higher CRPS of the IMERG climatological forecast, and the later period appeared, on average, to be slightly more difficult to predict. However, the relative skill of each model is robust over the two test years.
}
\end{figure*}

\begin{figure*}
\centering
\includegraphics[width=\textwidth]{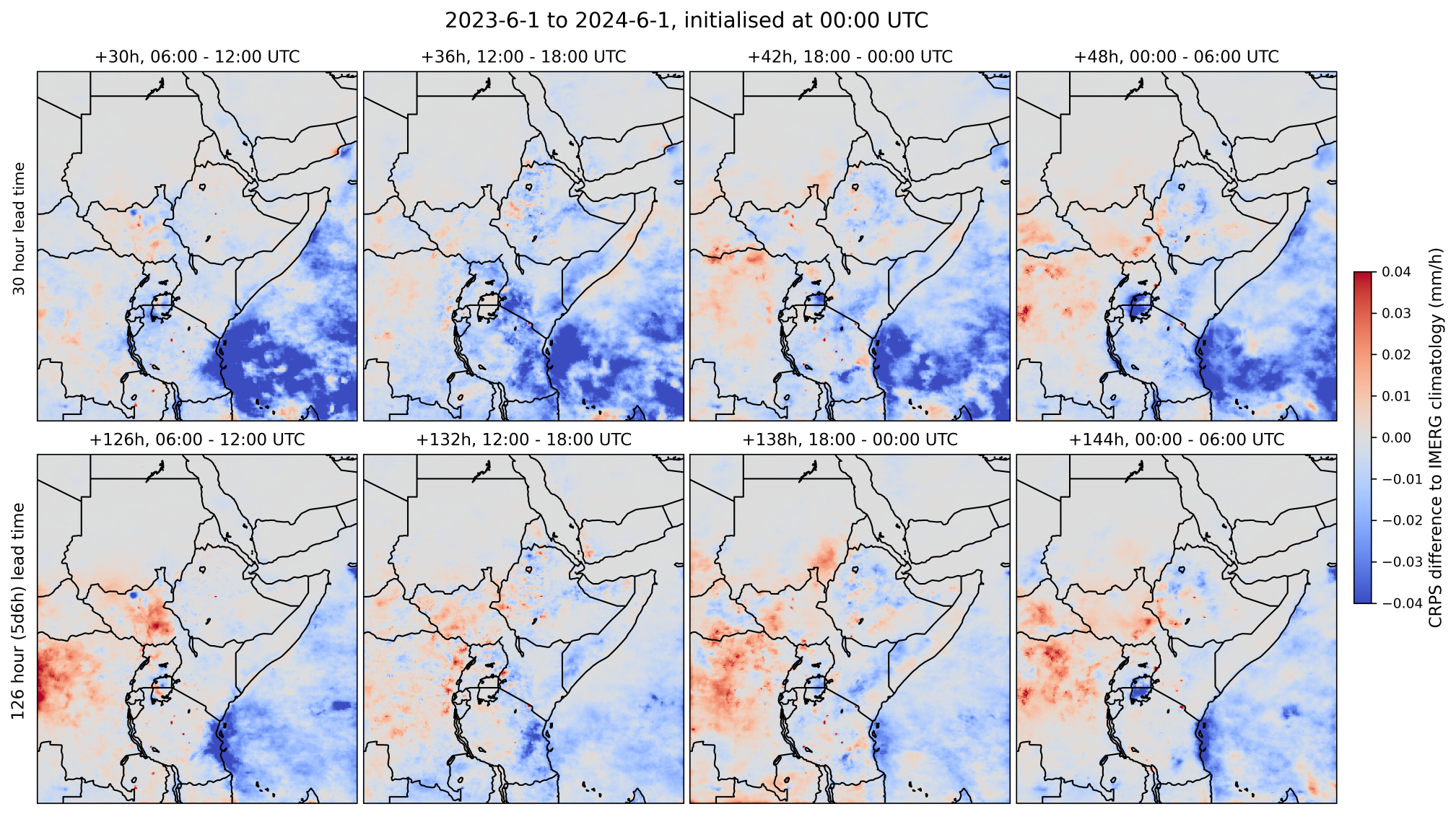}
\includegraphics[width=\textwidth]{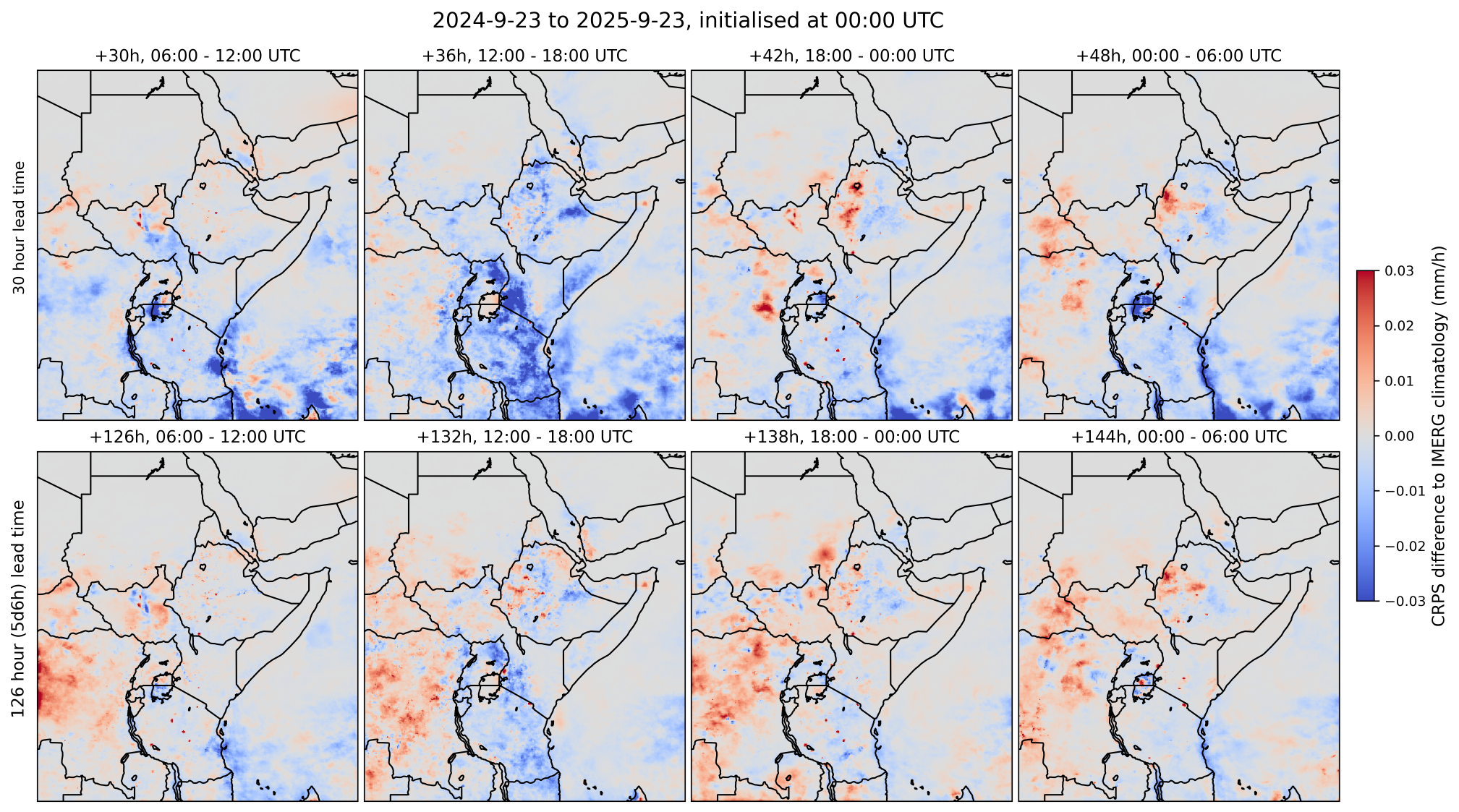}
\caption*{Figure S5: Difference between the one-year mean CRPS of the 6h rainfall accumulation cGAN forecasts and the CRPS of the IMERG climatological distribution at some example lead times. {\bf Top two rows:} the test period 2023-6-1 to 2024-6-1. {\bf Bottom two rows:} the test period 2024-9-23 to 2025-9-23. Note the change in colour scale. The lead time is given in hours. {\bf Row 1 and 3 from top:} Lead time starting 1 day 6 hours after initialisation at 00:00. {\bf Row 2 and 4 from top:} Lead time starting 5 days 6 hours after initialisation at 00:00. Blue means that cGAN has a lower (better) CRPS. Red means that the IMERG climatological forecast has a lower CRPS.}
\end{figure*}

\begin{figure*}
\centering
\includegraphics[width=0.7\textwidth]{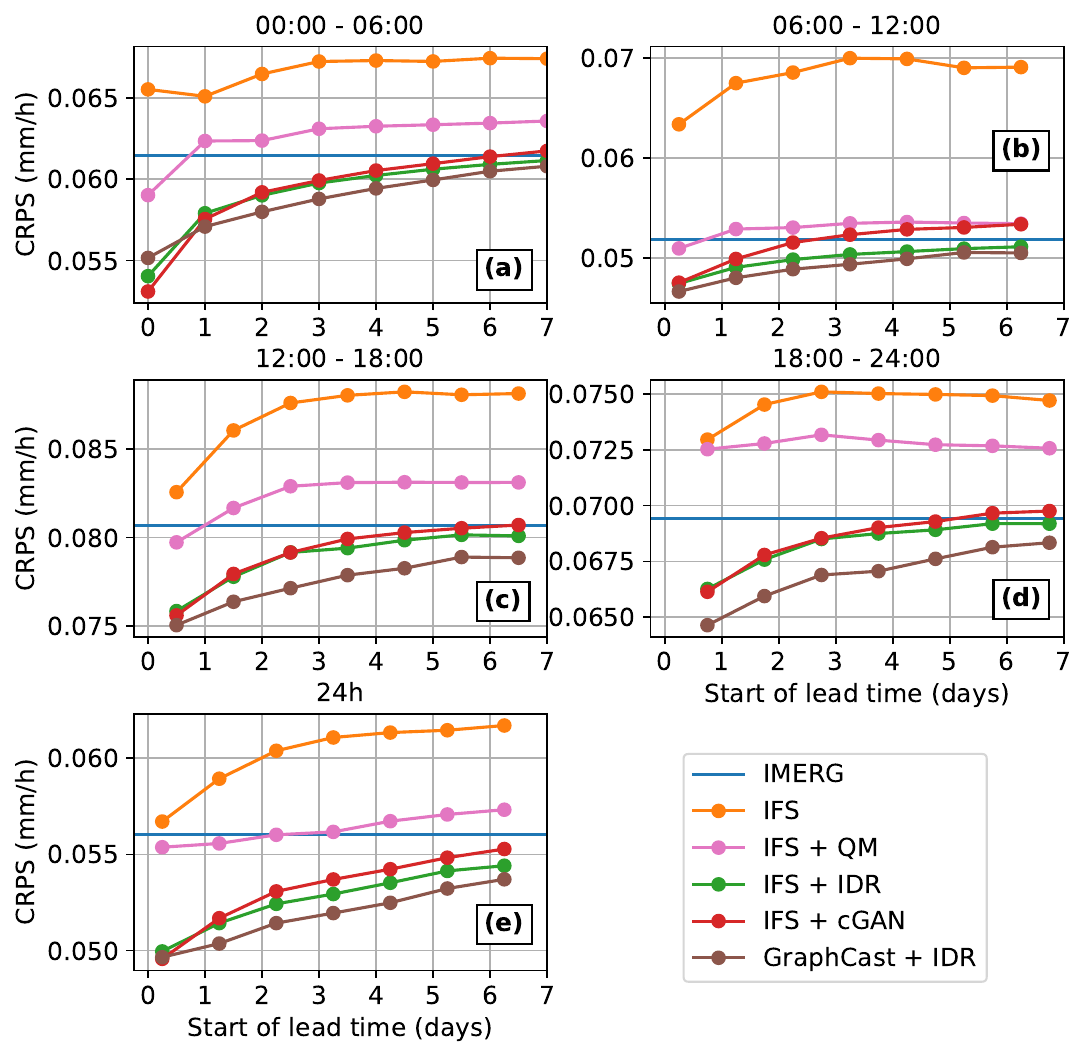}
\caption*{Figure S6: 6h and 24h rainfall accumulation CRPS values averaged over the East African domain and the one-year test period 2024-9-23 to 2025-9-23. Compare to Fig. 4 in the main paper. The top four plots (a,b,c,d) represent the four different 6 hour rainfall accumulation periods per day. Times are UTC and all forecasts are initialised at 00:00 UTC. The bottom plot (e) represents rainfall accumulation from 06:00 to 06:00. Each point represents the start of the accumulation period. QM stands for quantile mapping applied to each of 50 ensemble members. 1000 ensemble members are generated for the IFS+cGAN points. The blue line is the CRPS of the IMERG climatological distribution. Lower is better, but the domain average masks important issues, see Fig. 3.}
\end{figure*}

\begin{figure*}
\centering
\includegraphics[width=0.6\textwidth]{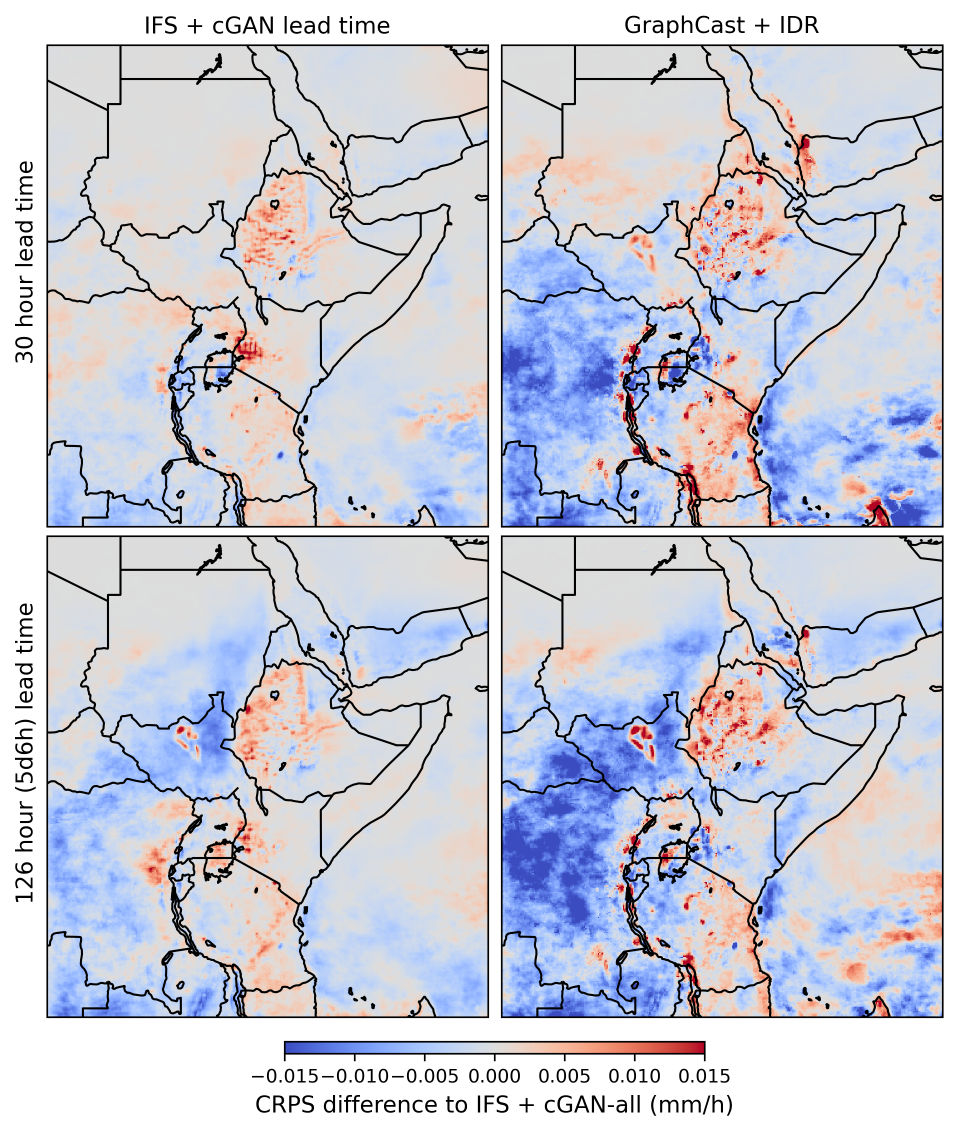}
\caption*{Figure S7: Difference between the one-year mean CRPS, over the period 2024-9-23 to 2025-9-23, of the 24h rainfall accumulation of 1000 ensemble members of the IFS+cGAN forecast trained on data from all forecast lead times, denoted {\it IFS+cGAN-all} and {\bf left column}: IFS+cGAN trained on data only from the lead time plotted denoted {\it IFS+cGAN lead time} and {\bf right column}: single ensemble member (deterministic) GraphCast with IDR post-processing to obtain a distribution. Blue means that the labelled model has a lower (better) CRPS. Red means that the IFS+cGAN-all forecast has a lower CRPS. Compare to Fig. 3 in the main paper.}
\end{figure*}

\begin{figure*}
\centering
\includegraphics[width=0.7\textwidth]{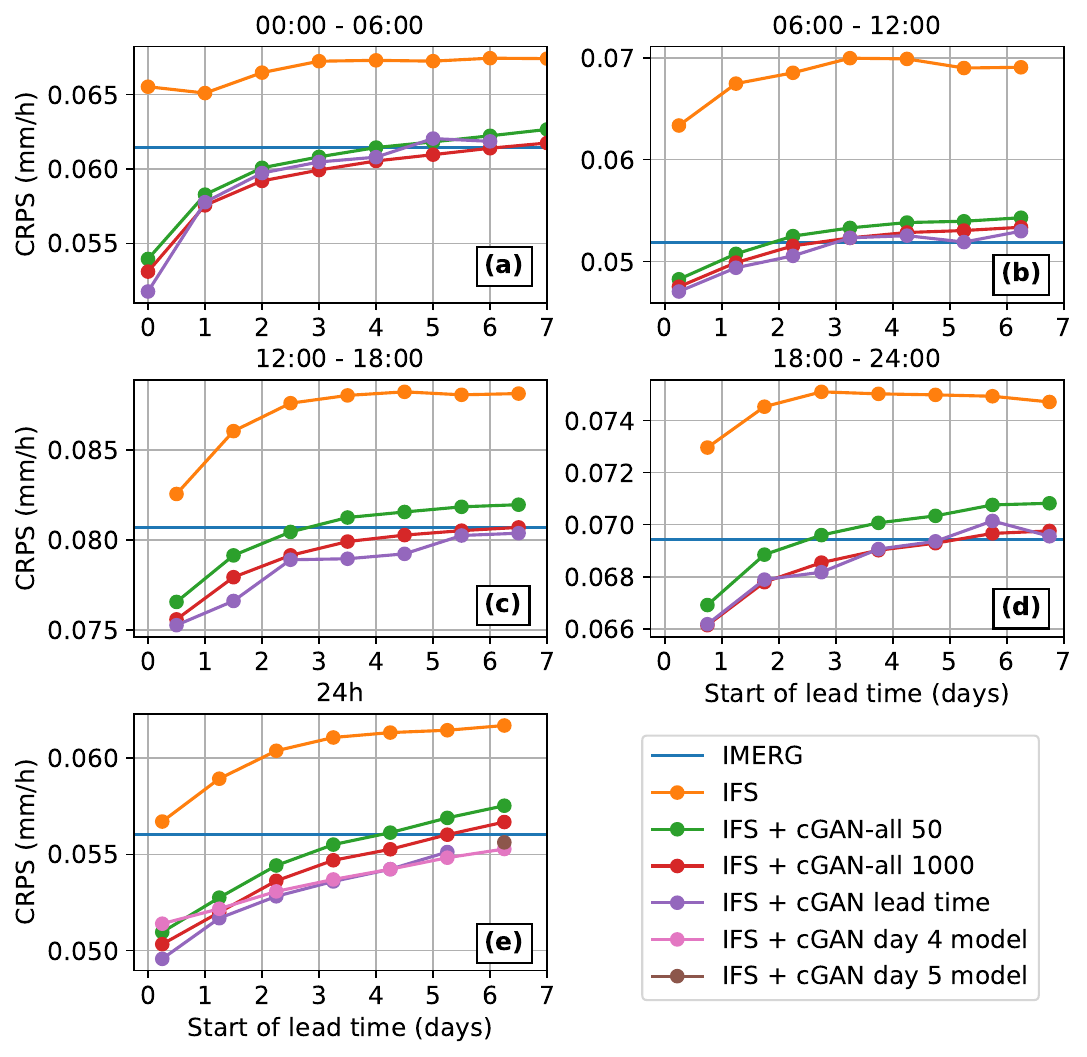}
\caption*{Figure S8: 6h and 24h rainfall accumulation, area and one-year time mean CRPS over the period 2024-9-23 to 2025-9-23. Compare to Fig. 6 in the main paper.
The top four plots (a,b,c,d) represent the four different 6 hour rainfall accumulation periods per day. Times are UTC. The bottom plot (e) represents rainfall accumulation from 06:00 to 06:00. All forecasts are initialised at 00:00 UTC. Each point represents the start of the accumulation period. The blue line is the CRPS of the IMERG climatological forecast. cGAN-all 50 denotes a 50 ensemble member forecast trained on all lead times. cGAN-all 1000 denotes the same but with 1000 ensemble members. cGAN lead time is the 1000 member forecast models trained separately on each lead time. In plot (e) the model trained on day 5 is used to forecast day 6 (brown point). The model trained on day 4 and applied to other days is also shown. Lower is better.}
\end{figure*}

\begin{figure*}
\centering
\includegraphics[width=0.6\textwidth]{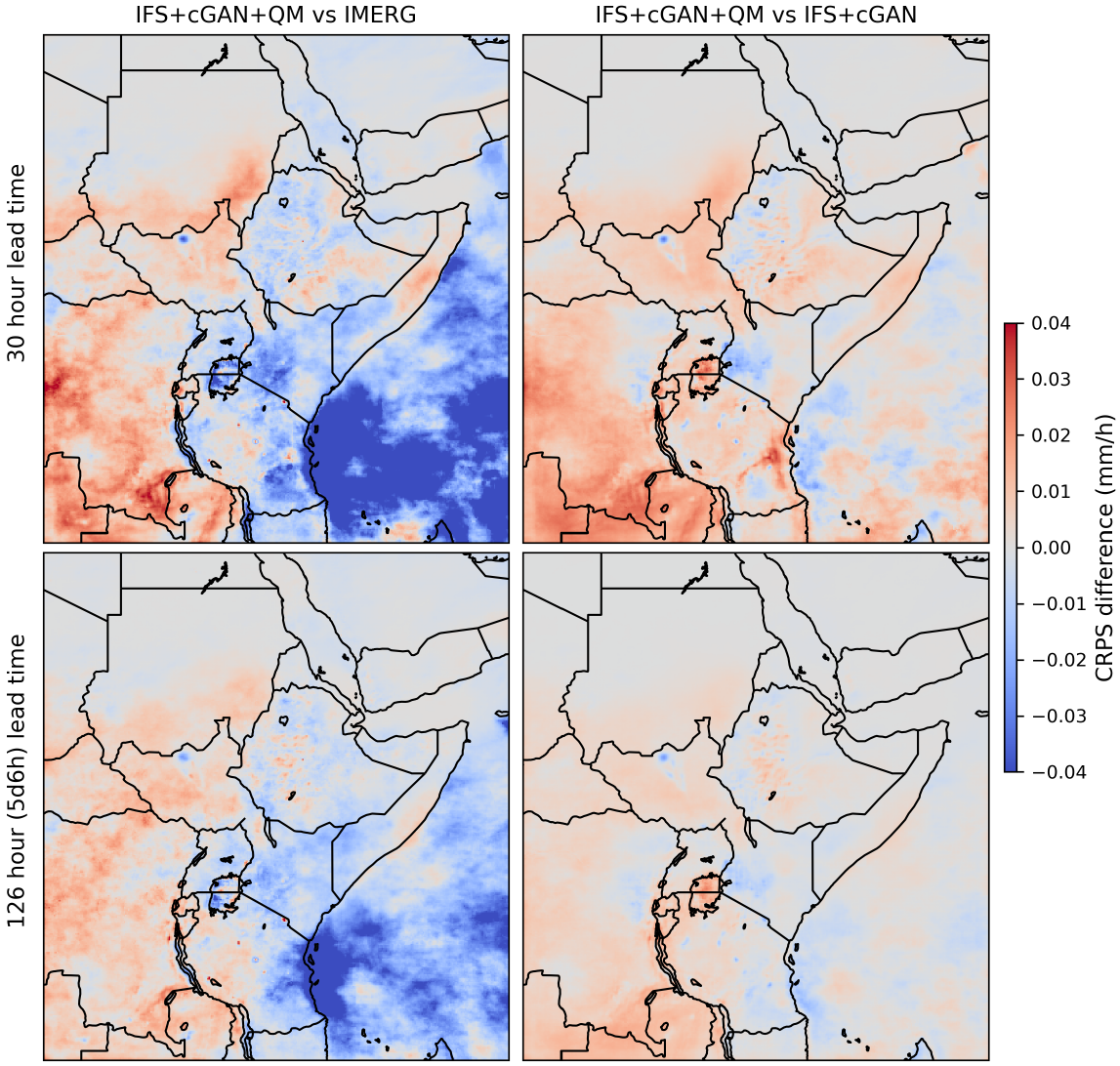}
\caption*{Figure S9: {\bf Left column:} Difference between the one-year mean CRPS of the 24h rainfall accumulation IFS + cGAN + quantile mapped forecast and the CRPS of the IMERG climatological distribution for {\bf top}, 1 day 6 hours lead time and {\bf bottom}, 5 days 6 hours lead time. Compare to Fig. 2. Blue means that cGAN has a lower (better) CRPS. Red means that the IMERG climatological forecast has a lower CRPS. {\bf Right column:} Same as the left plot but comparing to IFS + cGAN instead of IMERG. Although there are some areas of blue, particularly in south-western Kenya, large areas of the map are red indicating that the CRPS is higher (worse) when quantile mapping is applied.}
\end{figure*}

\begin{figure*}
\centering
\includegraphics[width=\textwidth]{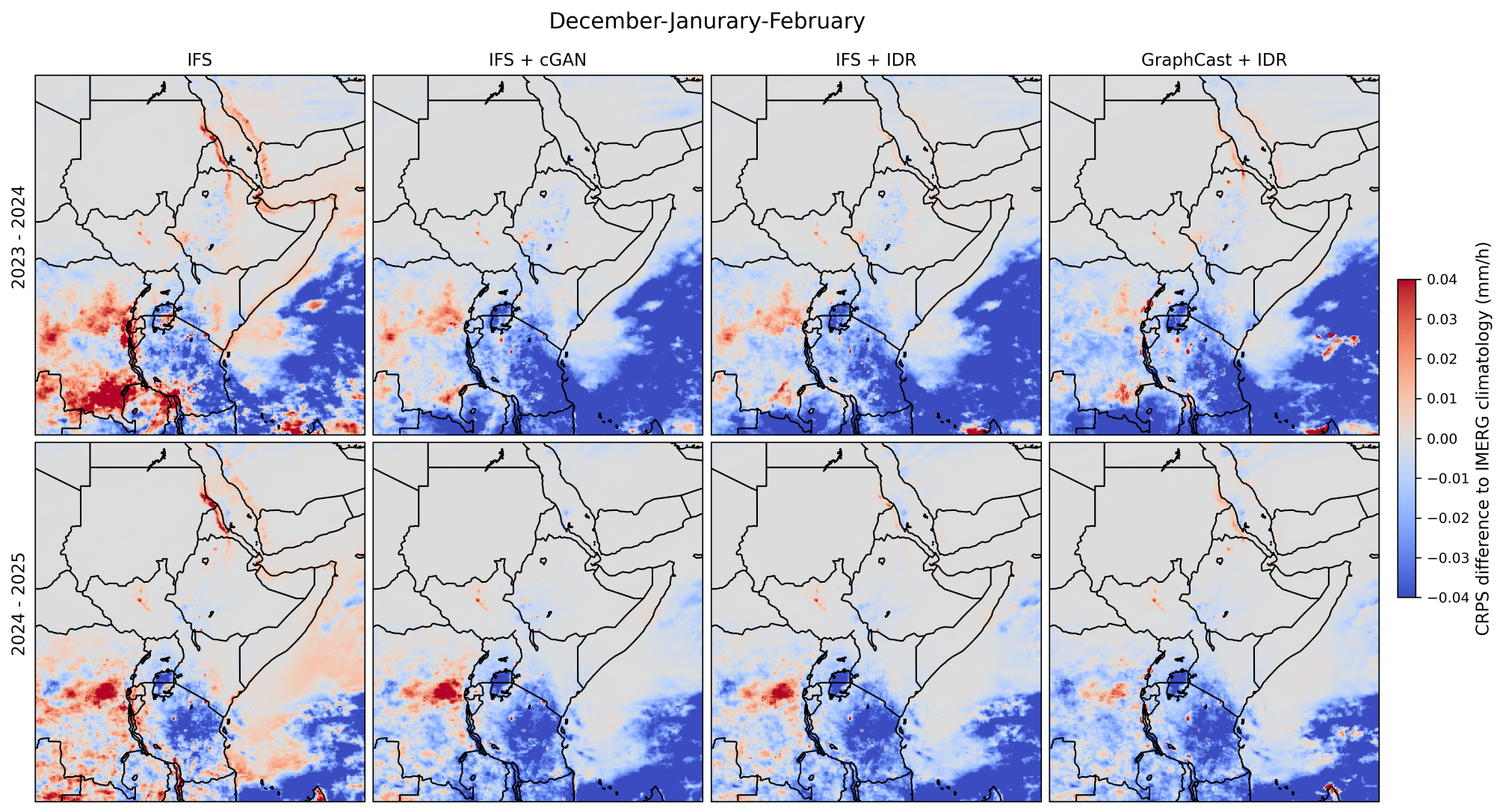}
\caption*{Figure S10: Difference between the CRPS of the 24h rainfall accumulation forecasts and the CRPS of the IMERG climatological distribution at a 30 hour lead time averaged over the December-January-February season in two example test periods, 2023-6-1 to 2024-6-1 (top) and 2024-9-23 to 2025-9-23 (bottom), for {\bf left}: IFS, {\bf middle left}: IFS with cGAN post-processing to 1000 ensemble members, {\bf middle right}: IFS with IDR post-processing and {\bf right}: GraphCast with IDR post-processing. Blue means that the model has a lower (better) CRPS. Red means that the IMERG climatological forecast has a lower CRPS. Compare to the amount of rainfall presented in Fig. S2.}
\end{figure*}

\begin{figure*}
\centering
\includegraphics[width=\textwidth]{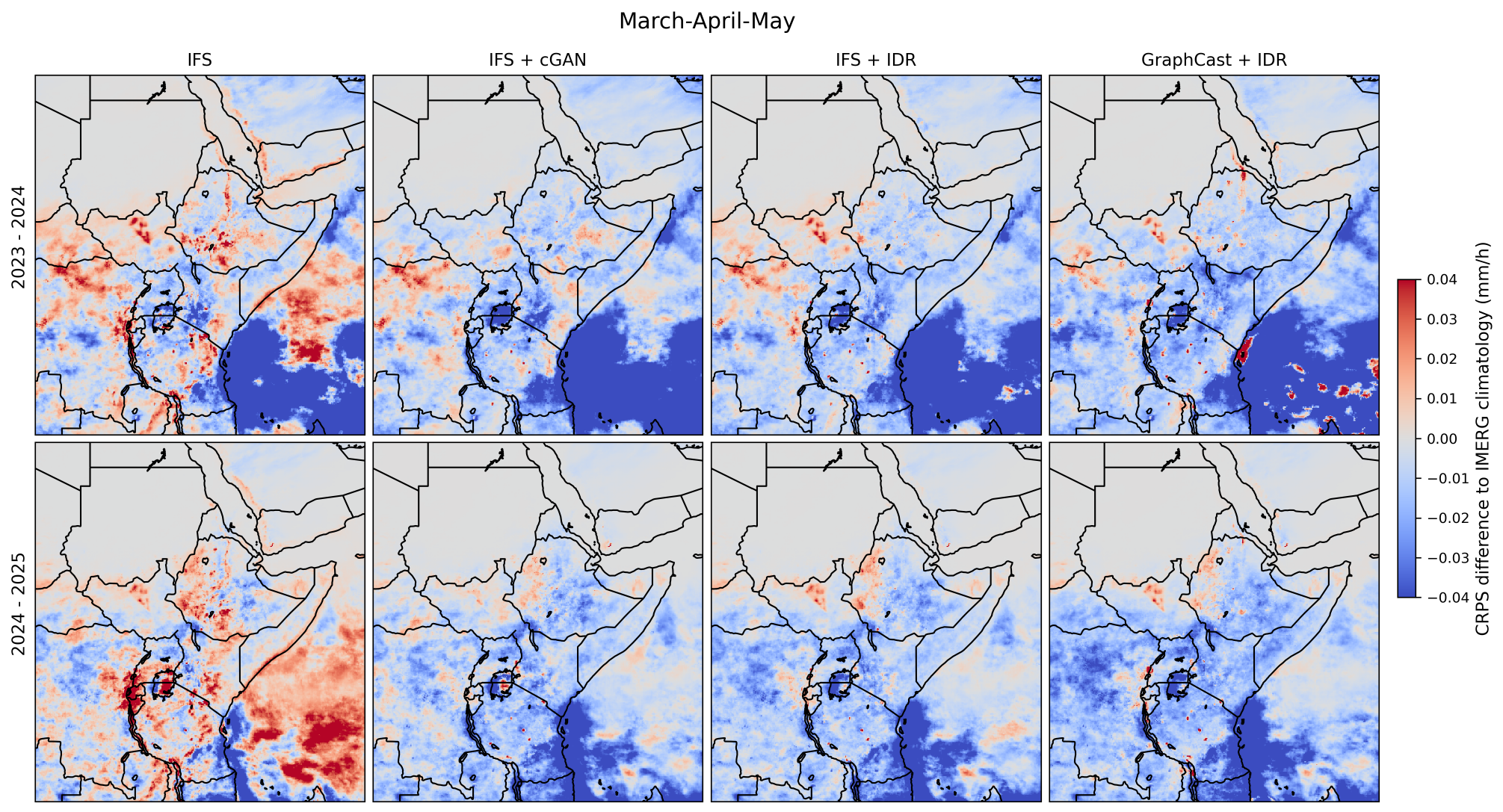}
\caption*{Figure S11: As Fig. S10 but for the March-April-May season. Difference between the CRPS of the 24h rainfall accumulation forecasts and the CRPS of the IMERG climatological distribution at a 30 hour lead time averaged over the March-April-May season in two example test periods, 2023-6-1 to 2024-6-1 (top) and 2024-9-23 to 2025-9-23 (bottom), for {\bf left}: IFS, {\bf middle left}: IFS with cGAN post-processing to 1000 ensemble members, {\bf middle right}: IFS with IDR post-processing and {\bf right}: GraphCast with IDR post-processing. Blue means that the model has a lower (better) CRPS. Red means that the IMERG climatological forecast has a lower CRPS. Compare to the amount of rainfall presented in Fig. S2.}
\end{figure*}

\begin{figure*}
\centering
\includegraphics[width=\textwidth]{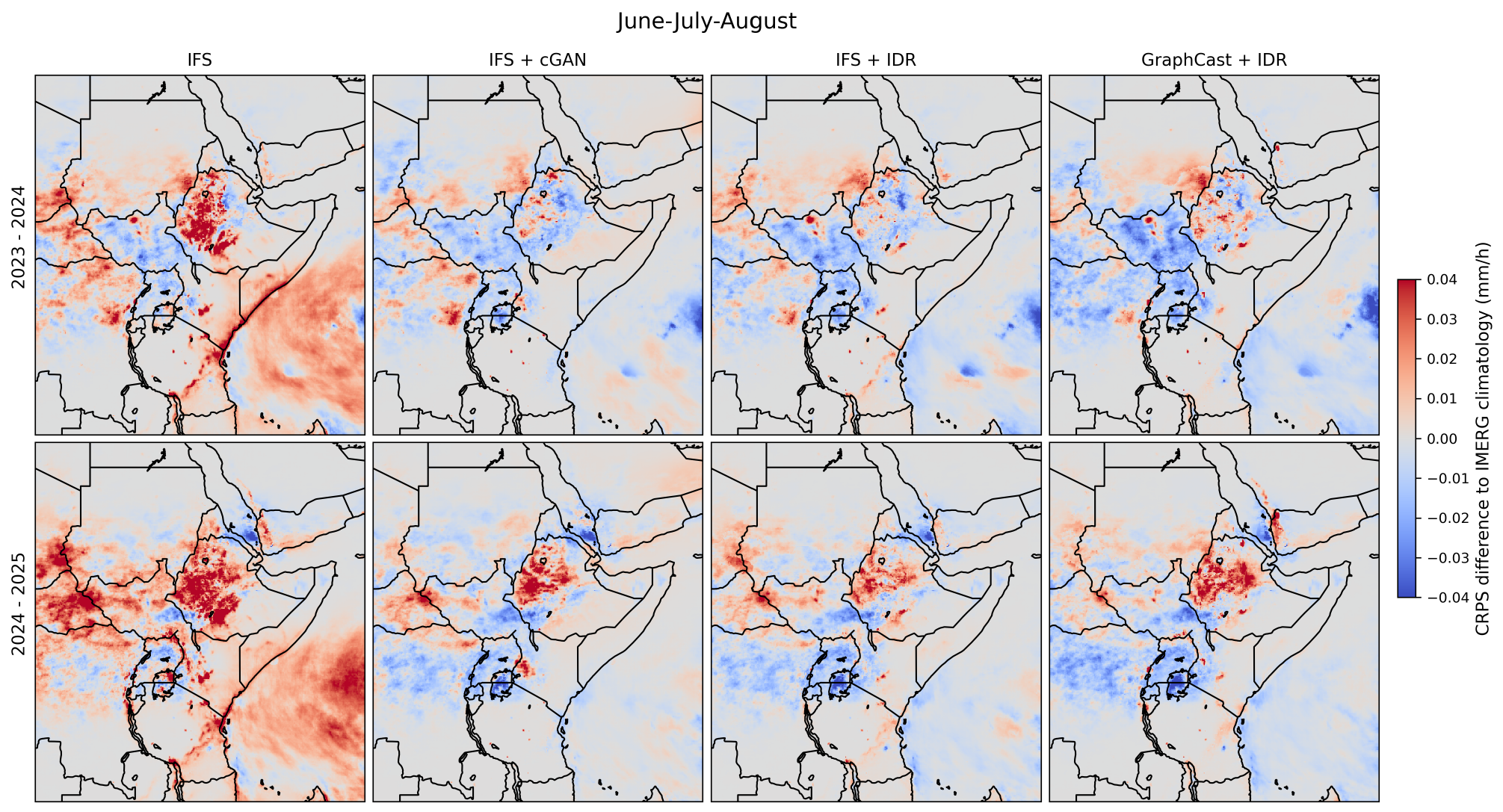}
\caption*{Figure S12: As Fig. S10 but for the June-July-August season. Difference between the CRPS of the 24h rainfall accumulation forecasts and the CRPS of the IMERG climatological distribution at a 30 hour lead time averaged over the June-July-August season in two example test periods, 2023-6-1 to 2024-6-1 (top) and 2024-9-23 to 2025-9-23 (bottom), for {\bf left}: IFS, {\bf middle left}: IFS with cGAN post-processing to 1000 ensemble members, {\bf middle right}: IFS with IDR post-processing and {\bf right}: GraphCast with IDR post-processing. Blue means that the model has a lower (better) CRPS. Red means that the IMERG climatological forecast has a lower CRPS. Compare to the amount of rainfall presented in Fig. S2.}
\end{figure*}

\begin{figure*}
\centering
\includegraphics[width=\textwidth]{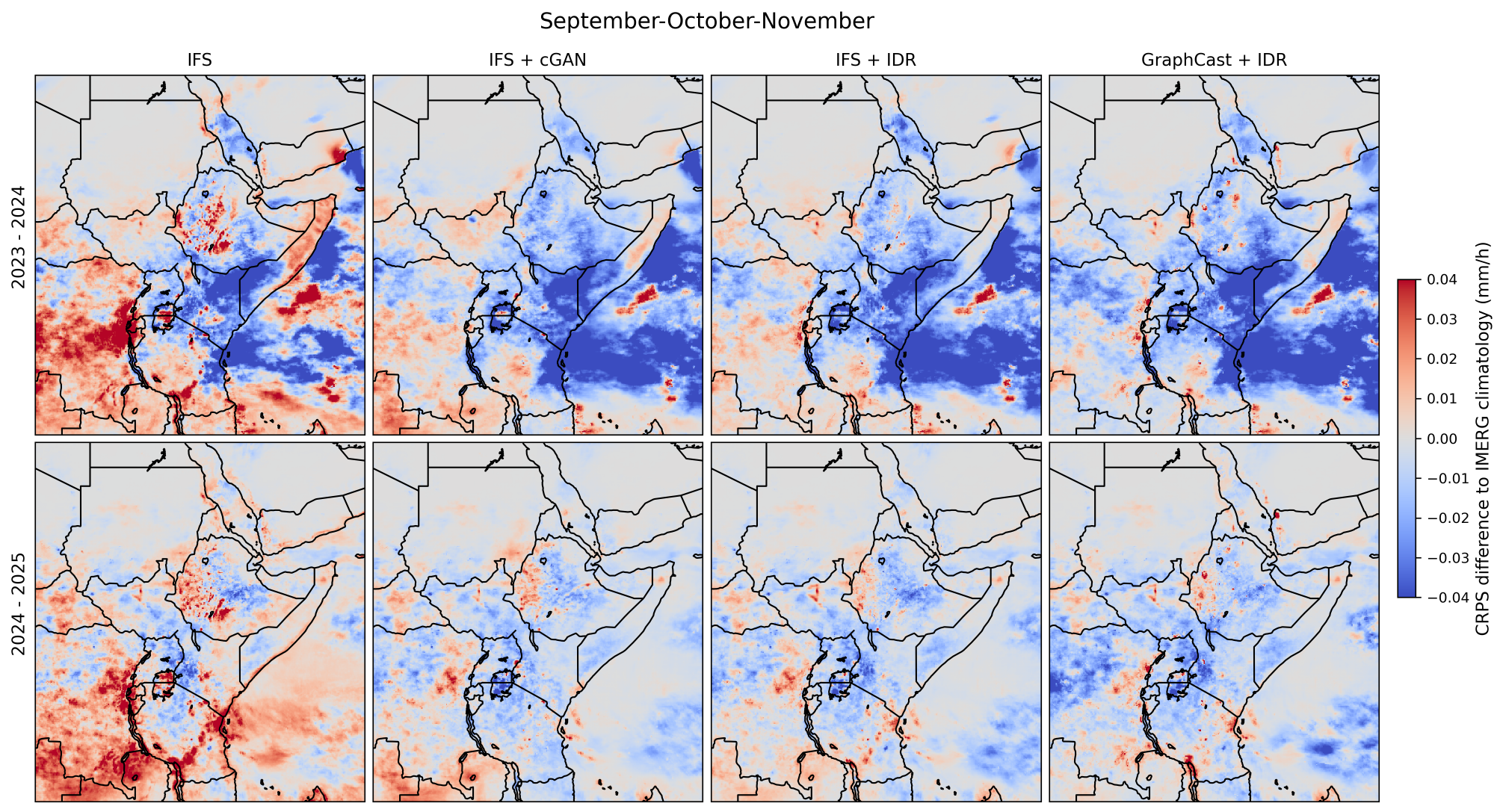}
\caption*{Figure S13: As Fig. S10 but for the September-October-November season. Difference between the CRPS of the 24h rainfall accumulation forecasts and the CRPS of the IMERG climatological distribution at a 30 hour lead time averaged over the September-October-November season in two example test periods, 2023-6-1 to 2024-6-1 (top) and 2024-9-23 to 2025-9-23 (bottom), for {\bf left}: IFS, {\bf middle left}: IFS with cGAN post-processing to 1000 ensemble members, {\bf middle right}: IFS with IDR post-processing and {\bf right}: GraphCast with IDR post-processing. Blue means that the model has a lower (better) CRPS. Red means that the IMERG climatological forecast has a lower CRPS. Compare to the amount of rainfall presented in Fig. S2.}
\end{figure*}

\begin{figure*}
\centering
\includegraphics[width=0.725\textwidth]{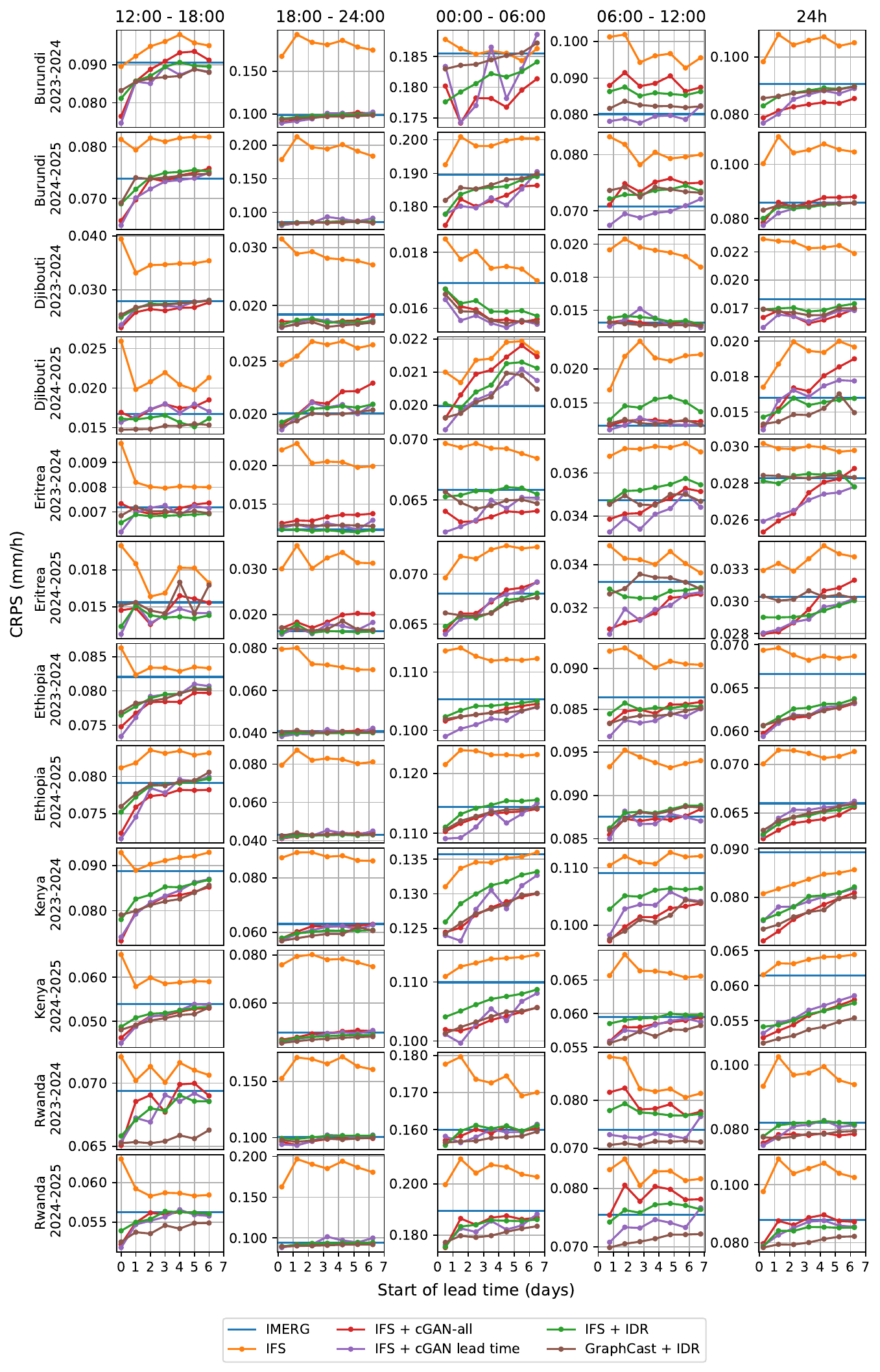}
\caption*{Figure S14: 6h and 24h rainfall accumulation CRPS averaged over each country and one year test period, as a function of lead time. Similar to Fig. 4 in the main paper but here the CRPS is masked to be computed only over individual countries. For ease of comparison, the results for the 2023-6-1 to 2024-6-1 test period are shown above the 2024-9-23 to 2025-9-23 test period. The first four columns represent the four different 6 hour rainfall accumulation periods per day. Times are UTC. The right column represents rainfall accumulation from 06:00 to 06:00. All forecasts are initialised at 00:00 UTC. Each point represents the start of the accumulation period. The blue line is the CRPS of the IMERG climatological forecast. cGAN all is the 1000 member forecast trained on all lead times at once. cGAN lead time is the 1000 member forecast models trained separately on each lead time. Lower is better, but country averages can still mask variations within each country, see Fig. 3 in the main paper. Continued in Fig. S15.}
\end{figure*}

\begin{figure*}
\centering
\includegraphics[width=0.725\textwidth]{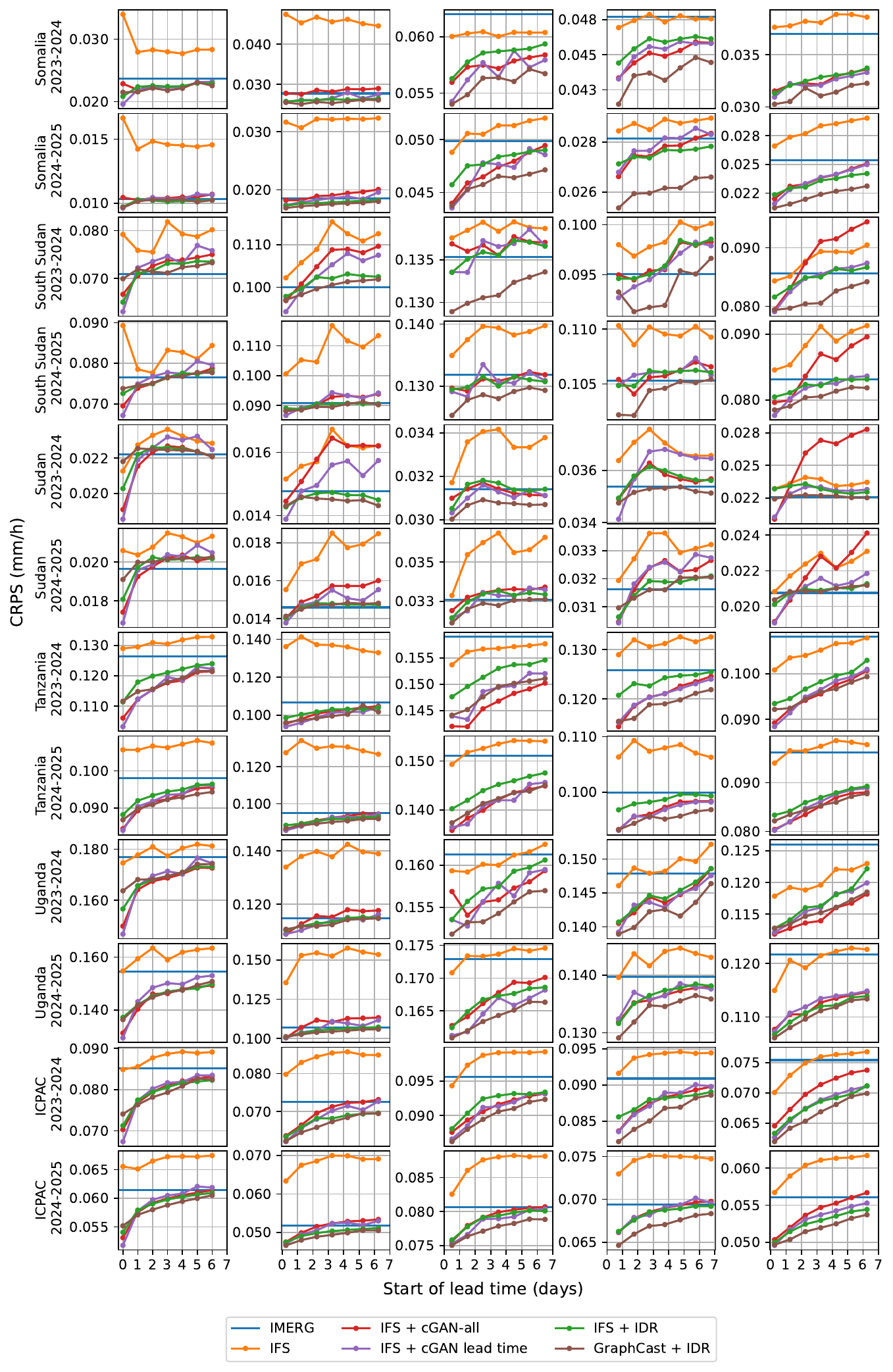}
\caption*{Figure S15: Figure S14 continued. 6h and 24h rainfall accumulation CRPS averaged over each country and one year test period, as a function of lead time. Similar to Fig. 4 in the main paper but here the CRPS is masked to be computed only over individual countries. For ease of comparison, the results for the 2023-6-1 to 2024-6-1 test period are shown above the 2024-9-23 to 2025-9-23 test period and the bottom two rows show curves that are also shown in Figs. 4, 6 and S6. The first four columns represent the four different 6 hour rainfall accumulation periods per day. Times are UTC. The right column represents rainfall accumulation from 06:00 to 06:00. All forecasts are initialised at 00:00 UTC. Each point represents the start of the accumulation period. The blue line is the CRPS of the IMERG climatological forecast. cGAN all is the 1000 member forecast trained on all lead times at once. cGAN lead time is the 1000 member forecast models trained separately on each lead time. Lower is better, but country averages can still mask variations within each country, see Fig. 3 in the main paper}
\end{figure*}

\begin{figure*}
\centering
\includegraphics[width=0.775\textwidth]{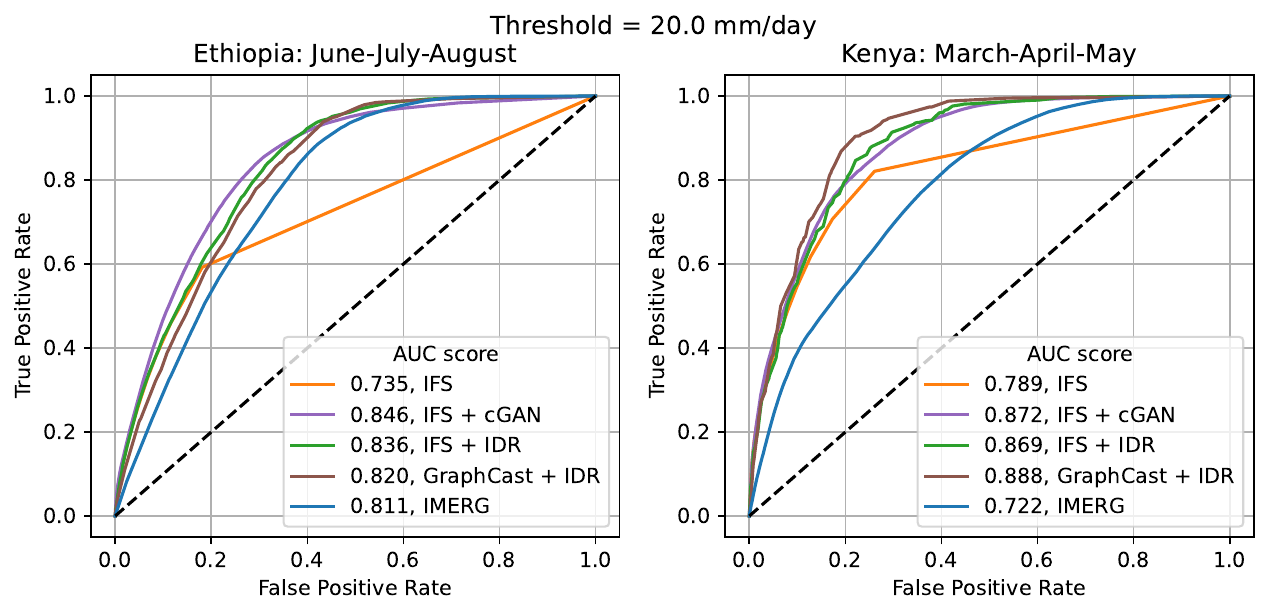}
\caption*{Figure S16: Example receiver operating characteristic (ROC) curves for the rainy seasons in {\bf left:} Ethiopia and {\bf right} Kenya. For a 20 mm/day rainfall threshold over the 2023-6-1 to 2024-6-1 test period, the true positive rate (the fraction of events where the forecast and the true rainfall both exceeded the 20mm/day threshold), is plotted against the false positive rate (the fraction of events where the forecast exceeded the 20mm/day threshold, but the true rainfall did not). The best forecast curve is when the monotonic curve hits the top left corner of the plot. A measure of the forecast quality is the area under the curve (AUC) score, with the best possible score being 1.0, see Figs. S17 and S18. In this example, the 24 hour rainfall accumulation period forecast begins with a lead time of 30 hours.}
\end{figure*}

\begin{figure*}
\centering
\includegraphics[width=0.775\textwidth]{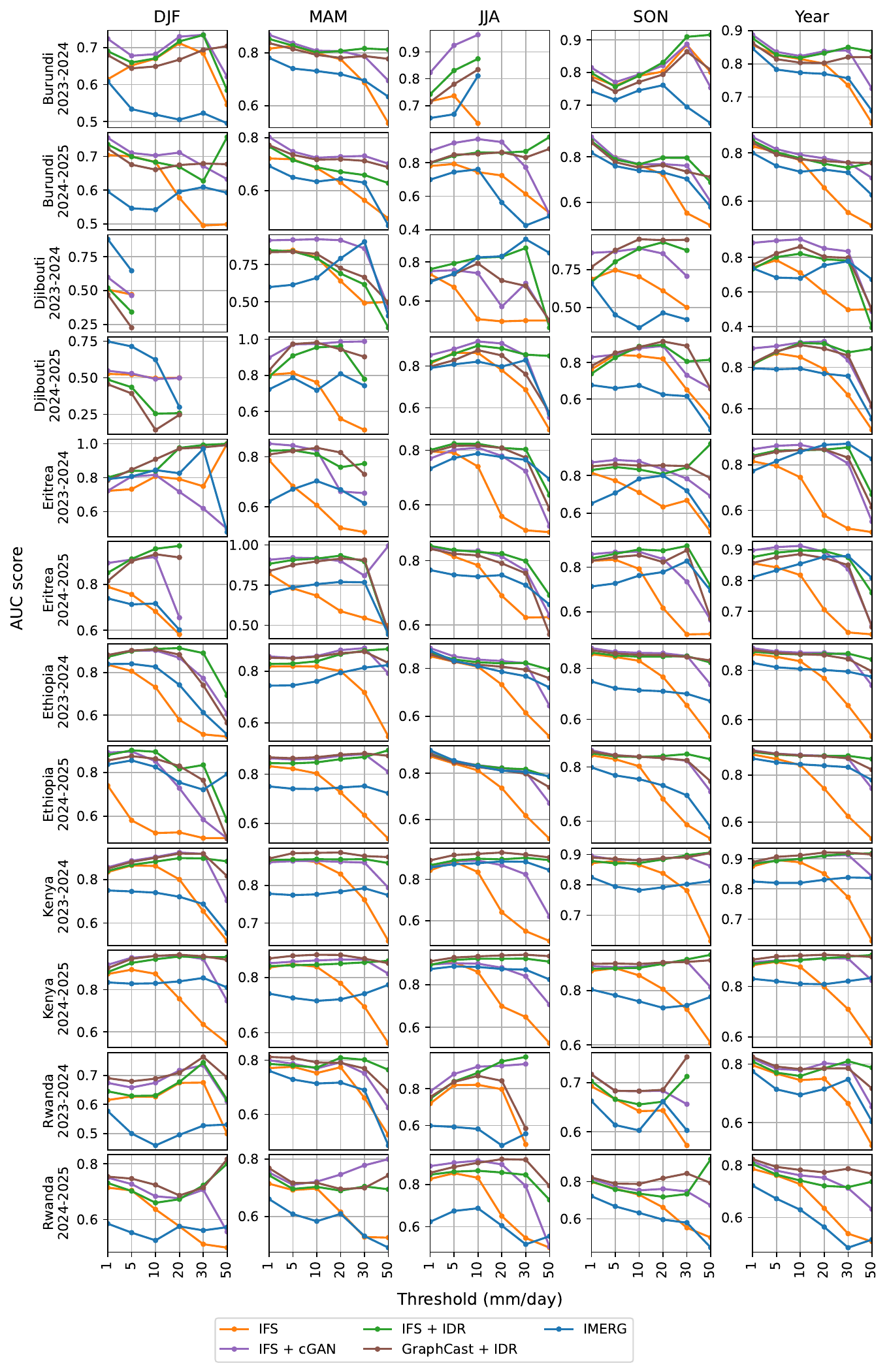}
\caption*{Figure S17: AUC scores derived from ROC curves for each season (first 4 columns) and the whole year (final column). See Fig. S16 for an example of the ROC curves. Scores are plotted for each country, continued in Fig. S18 along with the ICPAC region as a function of threshold. Missing points are where there is insufficient data that exceeds the given threshold. Both test years are shown to get an idea of the robustness of these results.}
\end{figure*}

\begin{figure*}
\centering
\includegraphics[width=0.775\textwidth]{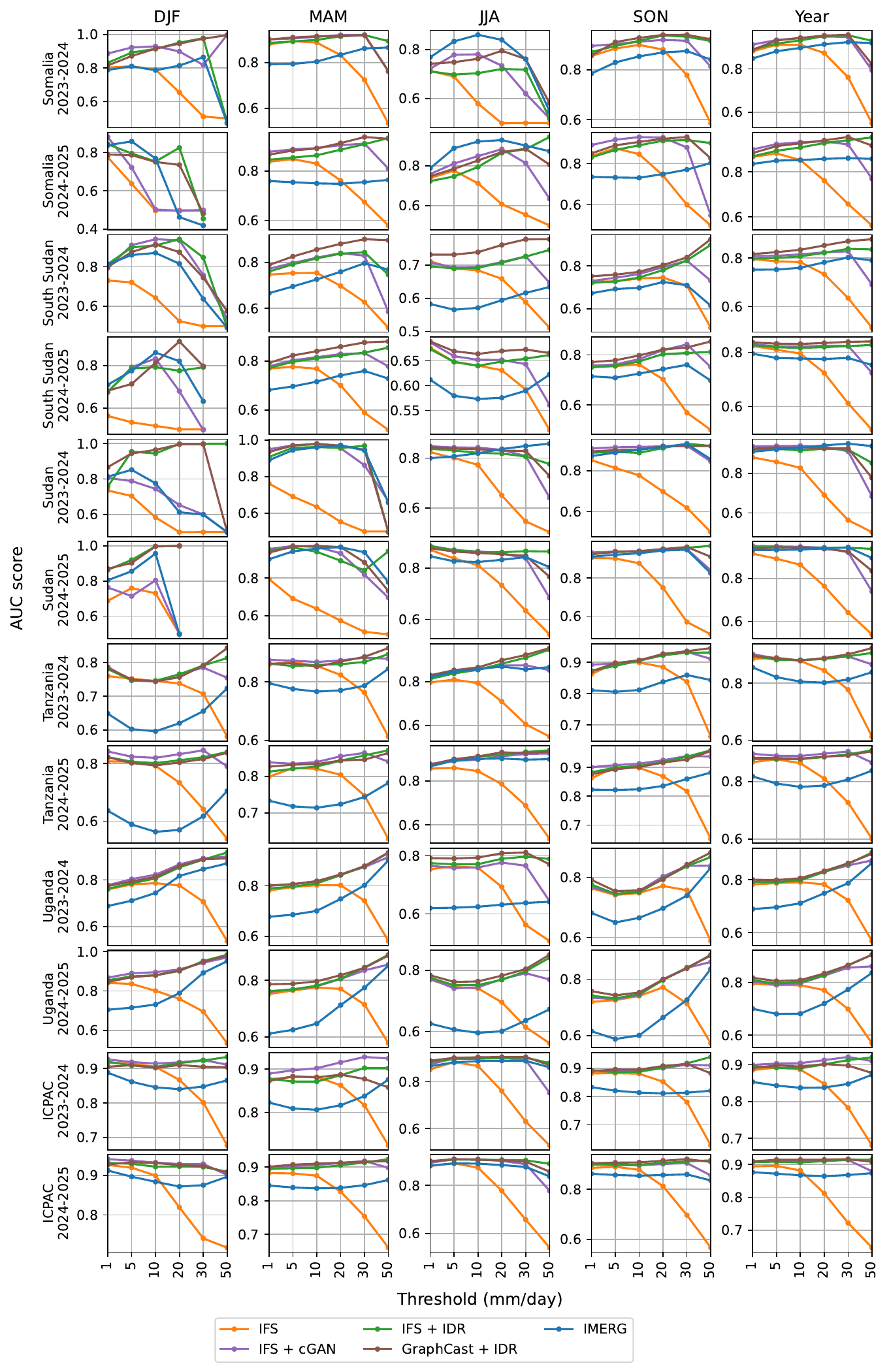}
\caption*{Figure S18: Continued from Fig. S17. AUC scores derived from ROC curves for each season (first 4 columns) and the whole year (final column). See Fig. S16 for an example of the ROC curves. Scores are plotted for each country along with the ICPAC region as a function of threshold. Missing points are where there is insufficient data that exceeds the given threshold. Both test years are shown to get an idea of the robustness of these results.}
\end{figure*}

\begin{table*}
\centering
\caption*{Table S1: Summary of models as used here and the reason for their inclusion. The newness of ECMWF's AIFS means that the data is not yet sufficient to train the post-processing models here. The IMERG v7 final data set stopped being produced in 2025, so there will also be a lack of the required overlap with AIFS operational forecasts.}
\begin{tabular}{|p{3.0cm}|p{7cm}|p{1.6cm}|p{3.5cm}|}
\hline
{\bf Model} & {\bf Details} & {\bf Ensemble size} & {\bf Figures} \\
\hline
IFS         & {\bf ECMWF's Integrated Forecast System (IFS).} A leading physics-based forecasting system, currently in operational use around the world.  & 50    & 2, 4, 5, 6, S3, S4, S6, S8, S10, S11, S12, S13, S14, S15, S16, S17, S18 \\ \hline
IFS + QM    & {\bf Quantile mapping applied to the IFS forecast.} Quantile mapping is a standard and well used rainfall post-processing technique. & 50    & 4, 5, S6 \\ \hline
IFS + IDR    & {\bf Isotonic Distributional Regression (IDR) applied to the IFS forecast.} The IDR is a recently introduced state-of-the-art statistical method that is gaining popularity. & Undefined    & 2, 4, S3, S4, S6, S10, S11, S12, S13, S14, S15, S16, S17, S18 \\ \hline
IFS + cGAN all    & {\bf The conditional Generative Adversarial Network (cGAN), a single model trained on all lead times and applied to the IFS forecast.} This is the model that is in daily use across East Africa. & 1000    & 3, 4, 5, 6, S6, S7, S8, S14, S15 \\ \hline
IFS + cGAN lead time    & {\bf The conditional Generative Adversarial Network (cGAN), multiple models each trained on a single lead time and applied to the IFS forecast.} If distribution biases vary across lead times, specialist models might show improved performance. & 1000    & 2, 3, 4, 6, S3, S5, S6, S7, S8, S9, S10, S11, S12, S13, S14, S15, S16, S17, S18 \\ \hline
IFS + cGAN + QM    & {\bf The IFS forecast post-processed with cGAN, then followed by Quantile Mapping.} To find out if the issues with the tails of the distribution in cGAN can be addressed with quantile mapping. & 1000    & 5, S9 \\ \hline
FuXi    & {\bf The FuXi pure machine learning forecast model.} One of the first competitive pure machine learning forecast models introduced. ECMWF produced daily forecasts using FuXi for evaluation, so evaluation data is available. & 1    & S4 \\ \hline
FuXi + IDR & {\bf The FuXi model post-processed using IDR.} To investigate if the lack of skill in FuXi is due to calibration of the forecast to ERA5 and lack of an ensemble, and to compare cGAN to another state-of-the-art statistical method. & Undefined & 2, 4, S4 \\ \hline
GraphCast    & {\bf The GraphCast pure machine learning forecast model.} Another one of the first competitive pure machine learning forecast models introduced. ECMWF produced daily forecasts using GraphCast for evaluation, so evaluation data is available. & 1    & S4 \\ \hline
GraphCast + IDR & {\bf The GraphCast model post-processed using IDR} To investigate if the lack of skill in GraphCast is due to calibration of the forecast to ERA5 and lack of an ensemble, and to compare cGAN to another state-of-the-art statistical method. & Undefined & 2, 3, 4, S3, S4, S6, S7, S10, S11, S12, S13, S14, S15, S16, S17, S18 \\ \hline
GenCast & {\bf The GenCast pure machine learning forecast model.} To compare a leading probabilistic pure machine learning model to IFS and put these results in more context. & 56 in 2023 52 in 2024 & S4 \\
\hline
\end{tabular}
\end{table*}

\end{document}